\input harvmac.tex

\def\neib{neighborhood}

\def\unlockat{\catcode`\@=11}
\def\lockat{\catcode`\@=12}

\unlockat

\def\newsec#1{\global\advance\secno
by1\message{(\the\secno. #1)}
\global\subsecno=0\global\subsubsecno=0\eqnres@t\noindent
{\bf\the\secno. #1}
\writetoca{{\secsym} {#1}}\par\nobreak\medskip\nobreak}
\global\newcount\subsecno \global\subsecno=0
\def\subsec#1{\global\advance\subsecno
by1\message{(\secsym\the\subsecno. #1)}
\ifnum\lastpenalty>9000\else\bigbreak\fi\global\subsubsecno=0
\noindent{\it\secsym\the\subsecno. #1}
\writetoca{\string\quad {\secsym\the\subsecno.} {#1}}
\par\nobreak\medskip\nobreak}
\global\newcount\subsubsecno \global\subsubsecno=0
\def\subsubsec#1{\global\advance\subsubsecno by1
\message{(\secsym\the\subsecno.\the\subsubsecno. #1)}
\ifnum\lastpenalty>9000\else\bigbreak\fi
\noindent\quad{\secsym\the\subsecno.\the\subsubsecno.}{#1}
\writetoca{\string\qquad{\secsym\the\subsecno.\the\subsubsecno.}{#1}}
\par\nobreak\medskip\nobreak}

\def\subsubseclab#1{\DefWarn#1\xdef
#1{\noexpand\hyperref{}{subsubsection}%
{\secsym\the\subsecno.\the\subsubsecno}%
{\secsym\the\subsecno.\the\subsubsecno}}%
\writedef{#1\leftbracket#1}\wrlabeL{#1=#1}}
\lockat

\def\IB{\relax\hbox{$\inbarmod\kern-.3em{\rm B}$}}
\def\IC{\relax\hbox{$\inbarmod\kern-.3em{\rm C}$}}
\def\ID{\relax\hbox{$\inbarmod\kern-.3em{\rm D}$}}
\def\IE{\relax\hbox{$\inbarmod\kern-.3em{\rm E}$}}
\def\IG{\relax\hbox{$\inbarmod\kern-.3em{\rm G}$}}
\def\IGa{\relax\hbox{${\rm I}\kern-.18em\Gamma$}}
\def\IF{\relax{\rm l\kern-.18em F}}
\def\IH{\relax{\rm I\kern-.18em H}}
\def\IK{\relax{\rm I\kern-.18em K}}
\def\IL{\relax{\rm I\kern-.18em L}}
\def\IP{\relax{\rm I\kern-.18em P}}
\def\IR{\relax{\rm I\kern-.18em R}}
\def\IZ{\relax\ifmmode\mathchoice
{\hbox{\cmss Z\kern-.4em Z}}{\hbox{\cmss Z\kern-.4em Z}}
{\lower.9pt\hbox{\cmsss Z\kern-.4em Z}}
{\lower1.2pt\hbox{\cmsss Z\kern-.4em Z}}\else{\cmss
Z\kern-.4em
Z}\fi}

\def\CD {{\cal D}}

\def\CF {{\cal F}}
\def\CG {{\cal G}}

\def\CL {{\cal L}}
\def\CM {{\cal M}}
\def\CN {{\cal N}}
\def\CO {{\cal O}}

\def\CS {{\cal S}}

\def\CU {{\cal U}}


\def\p{\partial}



\def\Tr{\rm Tr}

\font\manual=manfnt
\def\dbend{\lower3.5pt\hbox{\manual\char127}}

\def\half {{1\over 2}}


\def\lieg{{\underline{\bf g}}}



\font\cmss=cmss10 \font\cmsss=cmss10 at 7pt

\def\boxit#1{\vbox{\hrule\hbox{\vrule\kern8pt
\vbox{\hbox{\kern8pt}\hbox{\vbox{#1}}\hbox{\kern8pt}}
\kern8pt\vrule}\hrule}}
\def\mathboxit#1{\vbox{\hrule\hbox{\vrule\kern8pt\vbox{\kern8pt
\hbox{$\displaystyle #1$}\kern8pt}\kern8pt\vrule}\hrule}}


\def\inbarmod{\ \vrule height1.5ex width.4pt depth0pt}

\font\cmss=cmss10 \font\cmsss=cmss10 at 7pt


\def\a1{{\cal A}^{1,1}}

\def\so{{\rm SO}}

\def\su{{\rm SU}}
\def\sq{/ \relax{\kern-.27cm /}}

\def\ie{{\it i.e.}}
\def\cf{{\it cf}}

%

\lref\SeWi{N. Seiberg, E. Witten, ``Electric-Magnetic Duality,
Monopole Condensation, And Confinement in $N=2$ Supersymmetric
Yang-Mills Theory ''
Nucl. Phys. B426 (1994) 19-52 (and erratum - ibid. B430 (1994)
485-486 )\semi
``Monopoles, Duality and Chiral Symmetry Breaking in
N=2 Supersymmetric QCD'', hep-th/9408099,
Nucl. Phys. B431 (1994) 484-550.}
\lref\lerchesw{W. Lerche, ``Introduction to
Seiberg-Witten theory and
its stringy origins'', hep-th/9611190, Lectures
given at the Spring School and Workshop on String Theory,
Gauge
Theory
and Quantum Gravity, Trieste, Italy, 18-29 Mar 1996, to
the Workshop
on Gauge Theories, Applied Supersymmetry, and Quantum Gravity,
Leuven, Belgium, 10-14 Jul 1995, and to the NATO Advanced
Study
Institute: Les Houches Summer School on Theoretical Physics,
Session 64: Quantum Symmetries, Les Houches, France, 1
Aug - 8 Sep
1995.}

\lref\AandB{E. Witten, in ``Proceedings of the Conference
on Mirror Symmetry", MSRI (1991).}

\lref\asplouis{P. Aspinwall and J. Louis, ``On the
ubiquity of K3 fibrations in
string theory'', Phys. Lett. {\bf B369}, 233}
\lref\asplect{P. Aspinwall, ``K3 surfaces and string
duality'',
hep-th/9611137, Rutgers preprint RU-96-98,
lectures given at the 1996 TASI school on fields, strings, and
duality, Boulder, CO}
\lref\witvar{E. Witten, ``String theory dynamics in various
dimensions'',
hep-th/9503124, Nucl. Phys. {\bf 443}, 85}
\lref\kachruvafa{S. Kachru and C. Vafa, ``Exact results for
$\CN=2$ compactifications of heterotic strings'',
hep-th/9505105, Nucl. Phys. {\bf B450}, 69}
\lref\fhsv{S. Ferrara, J.A. Harvey, A. Strominger and C. Vafa,
``Second quantized mirror symmetry'', hep-th/9505162,
Nucl. Phys. {\bf B361}, 59}
\lref\edandjoe{J. Polchinski and E. Witten, ``Evidence
for heterotic-type I string duality'', hep-th/9510169,
Nucl. Phys. {\bf B460}, 525}
\lref\horwit{P. Horava and E. Witten, ``Heterotic and type I
string dynamics from eleven dimensions'', hep-th/9510209,
Nucl. Phys. {\bf B460}, 506}
\lref\multiplekthree{P.S. Aspinwall and M. Gross,
``Heterotic-heterotic string duality and multiple
K3 fibrations'', hep-th/9602118,
Phys. Lett. {\bf B382}, 81}
\lref\vafwit{C. Vafa and E. Witten, ``Dual string pairs with
$\CN = 1$ and $\CN = 2$ supersymmetry in four dimensions'',
hep-th/9507050, Harvard preprint HUTP-95-A023}
\lref\vafwitol{C. Vafa and E. Witten, ``A one-loop
test of string duality'', hep-th/9505053, Nucl. Phys.
{\bf B447}, 261}

\lref\gunasiertown{M. G\"{u}naydin, G. Sierra and P. Townsend,
``The geometry of $\CN = 2$ Maxwell-Einstein supergravity
and Jordan algebras'', Nucl. Phys. {\bf B242}, 244}
\lref\cadavid{A.C. Cadavid, A. Ceresole, R. D'Auria and
S. Ferrara, ``Eleven dimensional supergravity compactified
on Calabi-Yau threefolds'', hep-th/9506144,
Phys. Lett. {\bf B357}, 76}
\lref\papatown{G. Papadopoulos and P.K. Townsend,
``Compactification of $D=11$ supergravity on spaces
of exceptional holonomy'', hep-th/9506150,
Phys. Lett. {\bf B357}, 300}
\lref\townsend{P.~Townsend, ``Eleven-dimensional supermembrane
revisited'', hep-th/9501069, Phys. Lett. {\bf B350}, 184}

\lref\hmrtwo{J. A. Harvey and G. Moore, ``Five-brane
instantons and $R^{2}$ couplings in $\CN=4$ string theory'',
hep-th/9610237, U. Chicago preprint EFI-96-38}
\lref\hm{J.A.~ Harvey and G.~ Moore,
``Algebras, BPS States, and Strings,'' hep-th/9510182,
Nucl. Phys {\bf B463}, 315}
\lref\hmalg{J.A. Harvey and G. Moore, ``On the
algebras of BPS states'', hep-th/9609017, U. Chicago preprint
EFI-96-31}

\lref\joed{J. Polchinski, ``Dirichlet-branes and Ramond-Ramond
charges'',
hep-th/9510017, Phys.Rev.Lett. {\bf 75}, 4724}
\lref\smallinst{E. Witten, ``Small instantons in string
theory'',
hep-th/9511030, Nucl. Phys. {\bf B460}, 541}
\lref\vafhag{C. Vafa, ``Gas of D-branes and the
Hagedorn density
of BPS states'', hep-th/9511088, Nucl. Phys. {\bf
B463}, 413}
\lref\bsvtop{M. Bershadsky, V. Sadov and C. Vafa,
``D-branes and topological field theories'',
hep-th/9511222, Nucl.
Phys.
{\bf B463}, 420}
\lref\vafinst{C. Vafa, ``Instantons on D-branes'',
hep-th/9512078, Nucl. Phys {\bf B463}, 435}
\lref\dougone{M.R. Douglas, ``Branes within branes'',
hep-th/9512077, Rutgers preprint RU-95-92}
\lref\dougtwo{M.R. Douglas, ``Gauge fields and D-branes'',
hep-th/9604198, Rutgers preprint RU-96-24}

\lref\stromvaf{A. Strominger and C. Vafa, ``Microscopic
origin of the Bekenstein-Hawking entropy'', hep-th/9601029,
Phys. Lett. {\bf B379}, 99}

\lref\jeffandy{J.A. Harvey and A. Strominger, `` The heterotic
string is a soliton'', hep-th/9504047, Nucl. Phys. {\bf B449},
535; Erratum: ibid., {\bf B458}, 456}

\lref\beckstrom{K. Becker, M. Becker and A. Strominger
``Fivebranes,
membranes and nonperturbative string theory'', hep-th/9507158,
Nucl. Phys. {\bf B456}, 130}
\lref\phasemf{E. Witten, ``Phase transitions in M theory and F
theory'',
hep-th/9603150, Nucl. Phys. {\bf B471}, 195}

\lref\fth{C. Vafa, ``Evidence for F-theory'', hep-th/9602022,
Nucl. Phys. {\bf B469}, 403}
\lref\vafmor{D.R. Morrison and C. Vafa, ``Compactifications
of F-theory on Calabi-Yau threefolds -- I'', hep-th/9602114,
Nucl. Phys. {\bf B473, 72}\semi
``Compactifications of F-theory on Calabi-Yau threefolds
-- II'',
hep-th/9603161, Nucl. Phys. {\bf B476}, 437}

\lref\gipol{E.G. Gimon and J. Polchinski, ``Consistency
conditions
for orientifolds and D manifolds'', hep-th/9601038,
Phys. Rev. {\bf D54}, 1667}
\lref\lotsofolk{M. Berkooz, R.G. Leigh, J. Polchinski,
J.H. Schwarz, N. Seiberg and E. Witten, ``Anomalies,
dualities, and topology of $d=6$ $\CN=1$ superstring vacua'',
hep-th/9605184, Nucl. Phys. {\bf B475}, 115}
\lref\dmw{M.J. Duff, R. Minasian and E. Witten,
``Evidence for heterotic/heterotic duality'', hep-th/9601036,
Nucl. Phys. {\bf B465}, 413}

\lref\gmmm{
A.~ Gorsky,~ I.~ Krichever,~ A.~ Marshakov,~ A.~ Morozov,~ A.~
Mironov,~
``Integrability and
Seiberg-Witten Exact Solution'',
hep-th/9505035,  Phys.Lett.B355: 466-474, 1995}
\lref\emilnickone{E. Martinec and N. Warner, ``Integrable
systems and
supersymmetric gauge theory'', hep-th/9509161, Nucl.
Phys. {\bf
B459}, 97}
\lref\WitDonagi{
R.~ Donagi and E.~ Witten, ``Supersymmetric Yang-Mills
Theory and
Integrable Systems'', hep-th/9510101, Nucl.Phys.{\bf
B460}, 299}
\lref\ruijone{S.N.M. Ruijsenaars, ``Finite-dimensional soliton
systems'', in {\it Integrable and Super-Integrable Systems},
B.A. Kuperschmidt, ed., World Scientific, Singapore (1989)}
\lref\bruschi{M. Bruschi and O. Ragnisco, ``Lax
representations
and complete integrability for the relativistic periodic Toda
lattice'',
Phys. Lett. {\bf A134}, 365}
\lref\inoz{V.I. Inozemtsez, ``The finite Toda lattices'',
Comm. Math.
Phys.
{\bf 121}, 629}

\lref\kklmv{S. Kachru, A. Klemm, W. Lerche, P. Mayr and
C. Vafa,
``Non-perturbative results on the point particle limit of
$\CN = 2$
heterotic string compactifications'',  hep-th/9508155,
Nucl. Phys.
{\bf B459}, 537}
\lref\SWselfdual{A. Klemm, W. Lerche, P. Mayr and C. Vafa,
``Self-dual strings and $\CN = 2$ supersymmetric field
theory",
hep-th/9604034, Nucl. Phys. {\bf B477}, 746}
\lref\kkv{S. Katz, A. Klemm and C. Vafa, ``Geometric
engineering of
quantum field theories'', hep-th/9609239, U. Chicago preprint
EFI-96-39}

\lref\bsvegs{M. Bershadsky, V. Sadov and C. Vafa, 
``D strings on D manifolds'',
hep-th/9510225, Nucl. Phys. {\bf B463}, 398}
\lref\kpmegs{S. Katz, M.R. Plesser and D. Morrison,
``Enhanced gauge symmetry
in type II string theory'', hep-th/9601108, Nucl. Phys
{\bf B 477}, 105}
\lref\aspcy{P.S. Aspinwall, ``Enhanced gauge symmetries
and Calabi-Yau threefolds'', hep-th/9511171,
Phys.Lett.B371:231-237}

\lref\kenu{M.~ Kontsevich, ``Enumeration of rational
curves via torus
actions'', hep-th/9405035}
\lref\given{A.~ Givental, ``Equivariant Gromov-Witten
Invariants'',
alg-geom/9603021}
\lref\giventwo{A.~Givental, ``A mirror theorem for toric
complete
intersections'', alg-geom/9701016}
\lref\manintree{Yu. I. Manin, ``Generating functions in
algebraic
geometry and
sums over trees'', alg-geom/9407005}
\lref\bott{R. Bott, ``A residue formula for holomorphic vector
fields'', Jour. Diff. Geom.
{\bf 1}, 311}
\lref\atibott{M.F. Atiyah and R. Bott, ``The moment map and
equivariant cohomology'',
Topology, {\bf 23}, 1}
\lref\eghmath{T. Eguchi, P. B. Gilkey and A.J. Hanson,
``Gravitation,
gauge theories, and differential geometry'', Phys. Rep.
{\bf C}, 213}
\lref\itzptwo{C. Itzykson, ``Counting rational curves on
rational
surfaces'',
Int. J. Mod. Phys. {\bf B8}, 3703}

\lref\phases{E. Witten, ``Phases of N=2 Theories in Two
Dimensions",
{\it Nucl. Phys.} {\bf B403} (1993) 159, {\rm
hep-th/9301042}.}
\lref\agmbig{P. Aspinwall, B. Greene and D. Morrison
``Calabi-Yau
moduli space,
mirror manifolds, and space-time topology change in
string theory'',
hep-th/9309097, Nucl. Phys. {\bf B416}, 414}
\lref\hosonoone{S. Hosono, A. Klemm, S. Theisen 
and S.T. Yau, ``Mirror symmetry,
mirror map and applications to Calabi-Yau hypersurfaces'',
hep-th/9308122,
Comm. Math. Phys. {\bf 167}, 301}
\lref\hosonorev{S. Hosono, A. Klemm and S. Theisen,
``Lectures on mirror
symmetry'', hep-th/9403096, from the proceedings of the
3rd Baltic Student Seminar, Helsinki, 1993, ``Integrable
models and strings''}
\lref\morple{D. Morrison and M.R. Plesser,``Summing the
instantons: quantum
cohomology and mirror symmetry in toric varieties'',
hep-th/9412136, Nucl. Phys. {\bf B440}, 279}
\lref\katzone{S. Katz, in ``Essays on mirror symmetry'',
S.T. Yau, ed.}
\lref\aspmor{P.S. Aspinwall and D.R. Morrison,
``Topological field theory and rational curves'', 
hep-th/9110048, Comm. Math. Phys. {\bf 151}, 245}
\lref\mircoll{{\it Essays on Mirror Manifolds}, S.-T.  
Yau, ed., International
Press (Hong Kong), 1992\semi
{\it Essays on Mirror Manifolds II}, B.
Greene and S.-T. Yau, eds.,
International Press (Cambridge, MA), 1997}
\lref\greenecy{B. R. Greene, ``String theory on Calabi-Yau
manifolds'', hep-th/9702155, lectures
given at Theoretical Advanced Study Institute in
Elementary Particle
Physics (TASI 96): Fields, Strings, and Duality, Boulder,
CO, 2-28 Jun 1996. }
\lref\bathyperg{V. V. Batyrev and D. von Straten,
``Generalized
hypergeometric functions and rational curves on
Calabi-Yau complete
intersections in toric varieties'', alg-geom/930710,
Comm. Math. Phys. {\bf 168}, 493}
\lref\konthom{M. Kontsevich, ``Homological algebra of mirror
symmetry'', alg-geom/9411108, Proc. of the 1994
International Congress of Mathematicians, Birkh\"{a}user,  
1995}
\lref\batduke{V.V. Batyrev, ``Variations of the mixed Hodge
structure of affine hypersurfaces in algebraic tori'',
Duke Math. J. {\bf 69}, 349}
\lref\batdual{V.V. Batyrev, ``Dual polyhedra and
mirror symmetry for Calabi-Yau hypersurfaces in
toric varieties'', J. Alg. Geom. {\bf 3}, 493}

\lref\conif{A. Strominger, ``Massless black holes and
conifolds in string theory'',
hep-th/9504090, Nucl. Phys. {\bf B451}, 96\semi
B.R. Greene, D.R. Morrison and A. Strominger, ``Black hole
condensation and the
unification of string vacua'', hep-th/9504145, Nucl. Phys
{\bf B451}, 109}

\lref\seibfive{N.~Seiberg, ``Five-dimensional SUSY field
theories,
nontrivial fixed points, and string dynamics'',
hep-th/9608111,
Phys. Lett. {\bf B388}, 753}
\lref\seibjer{N.~Seiberg, lectures given at the 1996-1997
Jerusalem
Winter School}
\lref\seibmor{D.~ Morrison and N.~Seiberg, ``Extremal
transitions
and five-dimensional supersymmetric field theory'',
hep-th/9609070, Nucl. Phys. {\bf B483}, 229}
\lref\nikitafive{N. Nekrasov, ``Five dimensional gauge
theories and
relativistic integrable systems'', hep-th/9609219,
talk given at the 3rd International Conference on
Conformal Field
Theories
and Integrable Models, Chernogolovka, Russia, 23-30 June,
1996}
\lref\moganseib{D. R. Morrison, O. J. Ganor and N.
Seiberg, ``Branes,
Calabi-Yau spaces, and the toroidal compactification of
the $\CN$=1
six-dimensional ${\rm E}_{8}$ theory'', hep-th/9610251,
Nucl. Phys.
{\bf B487}, 93}
\lref\antfertay{I. Antoniadis, S. Ferrara and T.R. Taylor,
``$\CN=2$ heterotic superstring and its dual theory in five
dimensions'',
hep-th/9511108, Nucl. Phys. {\bf B460}, 489}
\lref\dkvdp{M.R. Douglas, S. Katz and C. Vafa,
``Small instantons, del Pezzo surfaces and type I' theory'',
hep-th/9609071, Harvard preprint HUTP-96-A042}
\lref\seibmorint{K. Intriligator, D.R. Morrison and N. Seiberg,
``Five-dimensional supersymmetric gauge theories 
and degenerations of Calabi-Yau spaces'', 
hep-th/9702198, Rutgers preprint RU-96-99,
IAS preprint IASSNS-HEP-96/112}

\lref\shenkshort{S.H.~Shenker, ``Another length scale in
string theory?'',
hep-th/9509132, Rutgers preprint RU-95-53}
\lref\dougshort{M.R. Douglas, D. Kabat, P. Pouliot and
S.H. Shenker,
``D-branes and short distances in string theory'',
hep-th/9608024,
Nucl. Phys. {\bf B485}, 85}

\lref\peskinft{M.E.~ Peskin and D.V.~ Schroeder, {{\it
An Introduction to Quantum Field Theory}}, Addison-Wesley
(Reading, MA), 1995}

\lref\klemmrev{A. Klemm, ``On the geometry behind 
$\CN = 2$ supersymmetric effective actions in 
four dimensions'',
hep-th/9705131, Trieste/Karpacz lectures}
\lref\holomrev{J. Louis and K. F\"{o}rger, ``Holomorphic
couplings in string theory'', hep-th/9611184, Trieste  
lectures}

\lref\witstrong{E. Witten, ``Strong coupling expansion
of Calabi-Yau compactifications'', hep-th/9602070,
Nucl. Phys. {\bf B471}, 135}
\lref\banksdine{T. Banks and M. Dine, ``Couplings and
scales in
strongly coupled
heterotic string theory'', hep-th/9605136, Nucl. Phys
{\bf B479}, 173}

\lref\newton{J.M.~Figueroa-O'Farrill, C.~K\"ohl and B.~
Spence, ``Supersymmetry
and the cohomology of (hyper)K\"ahler manifolds'',
hep-th/9705161}
\lref\verbit{M.~Verbitsky, ``On the action of a Lie
algebra $so(5)$
on the cohomology
of a hyperk\"ahler manifold'', Func. Anal. and Appl. {\bf
24}(2)
(1990) 70-71}
\lref\roslhyper{A.A.~ Rosly, unpublished}
\lref\lefshetz{P.~ Griffiths and J.~ Harris, ``Principles of
algebraic geometry'', WiIey 1978}

\lref\adhm{M.F. Atiyah, N.J. Hitchin, V.G. Drinfeld and  
Yu. I. Manin,
``Construction of instantons'', Phys. Lett. {\bf A65}, 185}
\lref\christ{N.H. Christ, E.J. Weinberg and N.K. Stanton,
``General self-dual Yang-Mills solutions'', Phys. Rev.  
{\bf D18},
2013}

\lref\lmns{A.~ Losev, G.~ Moore, N.~ Nekrasov, S.~
Shatashvili,
``Four-Dimensional Avatars of 2D RCFT,''
hep-th/9509151, Nuc. Phys. Proc. Suppl. {\bf 46}, 130
\semi
`` Chiral Lagrangians, Anomalies, Supersymmetry,
and Holomorphy'', hep-th/9606082,
Nucl. Phys. {\bf B 484}, 196}

\lref\wepromise{A. Lawrence and N. Nekrasov,
``Prepotentials, strings,
and torus actions'', to appear}
\lref\mnspromise{G.~ Moore, N.~ Nekrasov and  S.~Shatashvili,
``Hyperk\"ahler quotients,
Bethe Ans\"atz and Gauge Dynamics'', in preparation}

\lref\bfss{T. Banks, W. Fischler, S.H. Shenker and L.  
Susskind,
``M theory as a matrix model: a conjecture'', hep-th/9610043,
Phys. Rev. {\bf D55}, 5112}
\lref\berkroz{M. Berkooz and M. Rozali, ``String  
dualities from
matrix theory'', hep-th/9705175, Rutgers preprint RU-97-37}
\lref\rozali{M. Rozali, ``Matrix theory and U-duality
in seven dimensions'', hep-th/9702136, U. Texas preprint
UTTG-06-97}
\lref\berkrozseib{M. Berkooz, M. Rozali and N. Seiberg,
``Matrix description of M theory on $T^{4}$ and $T^{5}$'',
hep-th/9704089, U. Texas preprint UTTG-12-97 (Revised version)}

\lref\kmvbps{A. Klemm, P. Mayr and C. Vafa, ``BPS
states of exceptional non-critical strings'',
hep-th/9607139, CERN preprint CERN-TH/96-184,
Harvard preprint HUTP-96/A031}
\lref\lmwnoncrit{W. Lerche, P. Mayr and N. P. Warner,
``Noncritical strings,
del Pezzo singularities and Seiberg-Witten curves'',
hep-th/9612085}
\lref\mnwnoncrit{J.A. Minahan, D. Nemeschansky
and N.P. Warner, ``Investigating the BPS spectrum
of non-critical $E_{n}$ strings'', hep-th/9705237,
U. Southern. Calif. preprint USC-97/006,
Santa Barbara/IPT preprint NSF-ITP-97-055}

\Title{ \vbox{\baselineskip12pt\hbox{hep-th/9706025}
\hbox{HUTP-97/A023}\hbox{ITEP-TH-22/97}}}
{\vbox{
\centerline{Instanton sums and five-dimensional gauge
theories} }}
\bigskip
\bigskip
\centerline{Albion Lawrence$^{\dagger}$
and Nikita Nekrasov$^{\ddagger}$
}
\vskip 0.1cm
\centerline{$^{\dagger,\ddagger}$\it Lyman Laboratory of  
Physics, Harvard University,
Cambridge MA 02138, USA}
\centerline{$^{\ddagger}$\it Institute of
Theoretical and Experimental Physics,
117259, Moscow, Russia}
\vskip 0.3cm
\centerline{${}^{\dagger}$lawrence, ${}^{\ddagger}$nikita@string.harvard.edu}
\vskip 0.5cm
\centerline{\it}
\vskip 0.5cm
We analyze the vector multiplet prepotential of
$d=4$, $\CN=2$ type IIA compactifications.
We find that the
worldsheet instanton corrections
have a natural interpretation as one-loop corrections
in five dimensions, with the extra
dimension being compactified on a circle
of radius $g_{s}\ell_{s}$.
We argue that the relation between spacetime and
worldsheet instantons is natural from this point of view.
We also discuss the map between the type IIA worldsheet  
instantons
and the spacetime instantons in the heterotic dual.
\vskip 0.1cm
\Date{June 1997}

\newsec{Introduction}

Recent studies of non-perturbative string dynamics
have revealed that five dimensional gauge theories
contain rich and interesting physics.
This is not an academic issue, as
such theories arise in the following contexts:

\item{1.} Compactifications of M-theory on Calabi-Yau
threefolds \cadavid\papatown\phasemf;
\item{2.} The dynamics of D4-brane probes in type
I~$^{\prime}$ string theory \seibfive\seibmor;
\item{3.} Compactification of heterotic/type $I$ string
on $\bf K3 \times S^{1}$ \antfertay;
\item{4.} Chern-Simons theories for $\CN_{ws}=2$ strings
and $WZW_{4}$ models \lmns\nikitafive.

In addition, if string/M theory describes the world, then
the passage from five to four dimensions might teach us
something useful. For example, the 
${\rm E}_{8}\times{\rm E}_{8}$
heterotic string is dual to M-theory compactified on 
${\bf S}^{1}/{\IZ}_{2}$, with the heterotic string coupling 
proportional to the radius $\rho$
of the compactification \horwit.
For phenomenologically reasonable
$\CN = 1$ compactifications of the heterotic string on
a Calabi-Yau threefold with volume $\ell_{c}$, $\rho$
is an order of magnitude larger than both the
$\ell_{c}$ and the eleven-dimensional Planck
scale $\ell_{p,11}$ \witstrong\banksdine.

In this paper we will concentrate on the first item of
this list,
compactified further on a circle of radius $R$
(and discuss the third item as a dual description); this  
theory is of course type IIA string theory compactified on
the same Calabi-Yau manifold, with string coupling
$g_{s}=(R/\ell_{p,11})^{3/2}$ \witvar\townsend.
The five-dimensional theory will have eight supercharges,
the minimal ($\CN=1$) supersymmetry in five dimensions.

For Lagrangians at most quadratic in derivatives,
the vector multiplet dynamics of $\CN=1$ theories in five
dimensions are completely  determined by the 
prepotential $\CF$, a piecewise polynomial
function of the vevs of the scalars in the vector
multiplets.
For M-theory compactified down to $\IR^{5}$ on a Calabi-Yau
threefold, the vector multiplets correspond to the
sizes of the two-cycles (minus one for the graviphoton).
The five-dimensional gauge theory is
completely determined by the classical intersection ring
of the CY manifold; its prepotential is simply
\eqn\fivedcube{
	\CF \rightarrow \sum {1\over{6}} C_{ijk} T^{i}T^{j}T^{k}
}
in a given K\"{a}hler cone, where $C_{ijk}$ are the
intersection numbers of divisors ($i$,$j$,$k$) and
$T^{i,j,k}$ are the areas of the two-cycles  related
by Poincar\'{e} duality (in units of $\ell_{s}$.) 
The only nontrivial phenomena
which can happen
in this theory are flop transitions and
contractions of some submanifolds to zero size, leading
to either enhanced gauge symmetry or to exotic,
nontrivial infrared fixed points 
\phasemf\seibfive\seibmor\dkvdp\seibmorint.  In this paper, 
we will
concentrate on the contraction of divisors to rational curves;
the resulting singularities lead to enhanced gauge symmetry
without additional massless hypermultiplets 
\aspcy\kpmegs\phasemf\seibmor.

The prepotential $\CF$ becomes more interesting when we
compactify the five-dimensional theory on a circle
$\bf  S^{1}$ of radius $R$.
The prepotential of the resulting
four dimensional theory becomes a function of $R$,
such that in the limit $R \to \infty$
it reduces to the cubic form given in \fivedcube,
while in the $R \to 0$ limit it has much more complicated
form
$$
	\CF \sim {\half} T^{2} \left( {\rm log}
		\left({{T}\over{{\Lambda}}} \right)^{2} +
		\sum_{n=1}^{\infty} {{f_{n}}\over{n}}
		 \left({{\Lambda}\over{{T}}}\right)^{4n}  
\right)
$$
due to instanton corrections. Here $T$ denotes a complex
scalar of the vector multiplet in
four dimensions. In addition to the real five dimensional
scalar it contains a Wilson
loop around $\bf S^{1}$.

The purpose of this paper is to use this five-dimensional
point of view in order to understand how nonperturbative gauge
dynamics arises from nonperturbative worldsheet dynamics  
in type IIA string theory.
In \kkv\ the authors have shown how to ``geometrically
engineer'' $\CN=2$ gauge theories in four dimensions.
Enhanced nonabelian gauge symmetries 
arise
in IIA compactifications on Calabi-Yau
manifolds from fibrations of ADE singularities over rational curves
\aspcy (for type IIB compactifications this
was found by \bsvegs).  At a generic point in
the moduli space of $\CN=2$ gauge theories, the gauge group is
broken to its Cartan subalgebra \SeWi.\ (for a review 
of $\CN=2$ gauge theories and  
references see \lerchesw).
Thus, in order to study the gauge dynamics,
it suffices to compute the prepotential for the Abelian
gauge group
(with the caveat that we must also understand the  
singularities of the prepotential.)
The relevant vector multiplets come from the sizes of the
shrinking
$\IP^{1}$s in the ADE singularities.  Because the IIA
dilaton lives in a hypermultiplet,
the prepotential can be calculated exactly by a
tree-level string computation.

We would like to isolate the gauge theory part of the
string theory dynamics.  To do this, we must work in
a region of the moduli space of the theory in which the
relevant gauge theory scales -- the $W$ mass and the
QCD scale $\Lambda$ -- are much lighter than the
Planck mass or the string scale.  We can do this
by first taking $\ell_{s} \to \infty$.  The $W$ boson
will come from a Dirichlet two-brane wrapped around
a holomorphic two-cycle with area $a$; its mass
will be
$$
	M_{W} = {a \over g_{s} \ell_{s}^{3}} \  .
$$
Thus, in our problem we must have
$$
	a \ll g_{s} \ell_{s}^{2}\ .
$$
This means that for weak string coupling (a limit we will not
necessarily restrict
ourselves to), the size of the cycle must be smaller 
than the string scale.
This may be disturbing to some, but there are indications that
D-branes
probe geometry down to scales of order $g_{s}\ell_{s}$ in  
this limit \shenkshort\dougshort.
We also want to scale the K\"{a}hler classes so
that only those classes relevant to the ADE
fibration are relevant (\ie\ the other worldsheet instantons
decouple).
In cases where the singularity of interest can be  
imbedded in a
compact Calabi-Yau threefold, one can compute the full vector
multiplet prepotential and take the K\"{a}hler classes
one is not interested in to be
very large.  We can model this by working instead
with a noncompact Calabi-Yau threefold which is a fibration
of an ALE space over a two-sphere $\IP^{1}$.
In general, the prepotential can be expressed (within the
radius of convergence of the worldsheet instanton sum) as:
\eqn\fullinst{
	 \CF = {1\over{6}}C_{ijk}T_{i}T_{j}T_{k} +
\sum_{\{d_{i}\} \geq 0, k
\geq 1}
		{c_{\{d_{i}\}} \over k^{3}}
\prod_{i}e^{-d_{i}k T_{i}}\ ,
}
where $T_{i}$ are the K\"{a}hler classes in nonlinear
$\sigma$ model
coordinates,
$d_{i}$ specifies the degree of the ``primitive'' worldsheet
instantons,
and $k$ labels
the multiple coverings. (For a pedagogical review of
mirror symmetry
and Calabi-Yau compactifications, with further references,
see the lectures \greenecy.)  Clearly, upon
making the offending two-cycles large,
the related instanton contributions are
vanishingly small,
and the scalars determining their size
will haunt us only through the intersection numbers
$C_{ijk}$. (We will find that there are some subtleties in
working directly with the noncompact Calabi-Yau
manifolds; these scalars -- corresponding to the
Kaluza-Klein modes in the noncompact directions --
still appear in the cubic part of the prepotential and
are important when extracting the periods.)

Let us examine a specific model worked out in $\kkv$, namely
the local model of
an $\su(2)$ singularity (\ie\ an $A_{1}$ fibration).
We may model this by the total space $\CN(\IF_{0})$ of a
line bundle
 $\CL = \CO(-2)\otimes\CO(-2)$
over $\IF_{0} =\IP^{1}\times\IP^{1}$;
there is also a presentation of $\CN(\IF_{0})$ as a
noncompact toric variety,
which we will discuss in section 3 below.  The notation
$\CN(\IF_{0})$ comes from
the fact that the {\neib} of $\IF_{0}$ inside a compact
Calabi-Yau
manifold looks like its normal bundle, which in turn is  
isomorphic to $\CL$.
If we label one $\IP^{1}$ as the base ($b$) and the
other as the fiber ($f$), then in the limit that
gravity is turned off, the gauge coupling for the
$\CU(1)$ controlling the size of the
fiber is \kkv:
\eqn\fnoughtinst{
	\tau_{f} = i{\p^{2}\CF \over \partial T_{f}^{2}}
		= i \left( \sum_{n,m\geq 0; k>0} m^{2}
		{c_{n,m}\over k} q_{b}^{nk}q_{f}^{mk}\right) ,
}
where $q_{b} = e^{-T_{b}}$ and $q_{f} = e^{-T_{f}}$.
(With gravity present, the prepotential will be the same  
but the
gauge coupling will contain additional terms, also derivable
from the prepotential; these reflect
the difference between
special geometry and rigid special geometry.
See \holomrev\klemmrev\ for reviews and references.)
We would like to interpret this full instanton sum
in the context of the low-energy $\su(2)$ gauge theory.  Since
the $W$ bosons arise from Dirichlet
two-branes wrapped around the fiber, their mass is
proportional to $\vert T_{f} \vert$.  The
size of the base controls the gauge coupling:
$$
	T_{b} \approx {1 \over g^{2}}\ ,
$$
so that the enhanced symmetry point is $T_{f}\to 0$,
$T_{b}\to\infty$. (In fact, this identification
is a bit subtle, due to the aforementioned problems
with noncompact Calabi-Yau spaces.  We will discuss this
below). The authors of
\kkv\ show that if
this limit is approached as $T_{f}\sim \epsilon$,
$T_{b}\sim -4\log\epsilon$ for $\epsilon$ small, and
if we make the assumption that
\eqn\dnmconj{
	c_{n,m} \sim \gamma_{n}m^{4n-3}
}
for fixed $n$ and large $m$, then after
summing over $m$,
the leading ($\epsilon$-independent)
expression for the gauge coupling is:
$$
	\sum_{n}\gamma_{n} \left({e^{-T_b} \over
T_{f}^{4}}\right)^{n}\ .
$$
The appropriateness of this scaling limit can
be argued more rigorously as in \kklmv.
With the above expressions for $T_{b}$ and $T_{f}$ in
terms of gauge theory quantities, it
seems that we have reproduced the spacetime instanton
expansion of the
prepotential for the $\CN=2$ $\su(2)$ pure gauge theory
in $d=4$. One of the goals of this paper is to understand
how the relationship between the spacetime and
worldsheet instanton sums are realized dynamically in
string theory.

The bulk of this paper is an examination the sum
\fnoughtinst\ from the point of view
of the low energy gauge theory.  We will not take
the scaling limit discussed except when
we want to explicitly discuss the relation to
the work of Seiberg and Witten \SeWi.
The lesson we will learn will be that from the gauge theory point
of view the sum \fnoughtinst\ makes sense even away from
the scaling limit; it can be understood completely as
the one-loop correction to the gauge coupling of
a {\it five}-dimensional gauge theory compactified on
an additional circle of radius $g_{s}\ell_{s}$, with the
particle content specified by the string theory
compactification.
The existence of this extra dimension could be expected
given the relation of type IIA string theory to M-theory,
but the fact that the nonperturbative string theory
result \fnoughtinst\ can be rewritten as an entirely
perturbative field theory
calculation in five dimensions (for compactifications
with eight supersymmetries)
is interesting.

This paper will proceed as follows: in section 2
we will begin by reviewing
some facts about five-dimensional gauge theories compactified
on a circle.  We will then discuss how such theories
arise from
M-theory.  We will find that by summing over multiple
covers of primitive worldsheet instantons, the sum in equation
\fnoughtinst\ looks like a sum over one-loop terms
contributions
to the gauge theory prepotential in five dimensions.  We will
show that the Gromov-Witten invariants arise
naturally and physically
as the contributions of the spectrum of BPS states in the
M-theory compactification, weighted appropriately for
their spin.
We will then show that the natural ultraviolet scale in the
four-dimensional theory is in fact the ``Shenker scale''
$g_{s}\ell_{s}$ \shenkshort.  In section 3 we will
write down the expression for the periods determining the
prepotential
in terms of generalized hypergeometric functions.  We
will extract directly
a representation of these periods as integrals of a
one-differential
over the non-contractible  cycles in  Riemann surface, as
conjectured in \kkv.  We will
then show that the differential and Riemann surface have
a natural
interpretation in terms of the relativistic periodic Toda
lattice, as conjectured by one of the present authors in
\nikitafive.  We will also discuss the generalization to
$\su(n)$  gauge theories.  In section 4
we will state a conjecture about the relation
of the worldsheet instantons in compact Calabi-Yau models
to relevant instantons of strings and fivebranes in the
heterotic dual. Finally, we will present our conclusions and
wild speculations in section 5.

\newsec{Four- and five-dimensional gauge   theory
dynamics from type IIA/M-theory}

The minimal supersymmetric five-dimensional gauge theory
($\CN=1$) has eight
supercharges. It arises, for example,
from the dimensional reduction of the $\CN = 1$ theory in six
dimensions;
upon reduction to four dimensions it becomes an $\CN = 2$
theory.
(For discussions of five dimensional SUSY theories, see
\seibfive\seibmor\nikitafive\seibmorint.)  The
massless multiplets are
the hypermultiplet, with four real scalars and a
(four-component)
spinor; and the vector multiplet, with a real scalar, a
spinor,
and a vector (on shell). There is also a tensor multiplet,
which is electric-magnetic  dual to the vector multiplet.
We will also be interested in BPS
saturated states with arbitrary spin.  They will be classified
by the little group for five-dimensional massive particles,
\eqn\littlegroup{
	\so (4) = \su(2)_1 \times \su(2)_2 \ \ .
}

\subsec{One-loop corrections to the prepotential in
five dimensions}

We will conduct our analysis at a generic point on the vector
multiplet moduli space, where the gauge group $\CG$ is broken
to its Cartan subalgebra.  In the presence of
particles (for example, $W$ bosons)
with charges $e_{i}$ under the ${i}^{{\rm th}}$ $\CU(1)$
and with mass $M$,
the one-loop correction to the
gauge coupling
$$
	\tau_{i} = {4\pi i \over g_{i}^{2}}
$$
in five uncompactified dimensions is proportional to:
\eqn\fdol{
	i e_{i}^2 \int {{d^{5} k}\over{(k^{2} + M^{2})^{2}}} \sim
	e_{i}^2 \left( {\rm divergent} \ {\rm const.}
	+ i  | M | \right)
}
When we compactify to four dimensions on a circle
of radius $R$, there are finite size corrections to \fdol\
coming (in the first quantized picture) from worldline
instantons of the particle around this circle.
The worldline $k$-instanton action is
$2\pi k \vert M \vert R$, and the resulting correction to the
coupling is:
\eqn\instact{
	{i\over{2\pi R k}} e^{-2\pi R k | M |}\ .
}
Altogether these corrections sum up to:
\eqn\pert{
	\vert M \vert + 2 \sum_{k =1 }^{\infty}  
	{1\over{2\pi R k}}
	e^{-2\pi R k \vert M \vert} =
	{i\over{2\pi R}} {\rm log} \Bigl( {\rm sinh}^{2} ( \pi
	M R) \Bigr)\ .
}
One may also arrive at \pert\ by
writing the
momentum integral in \fdol\ as the product of a
four-dimensional
integral and a sum over the Kaluza-Klein momentum $p/R$
around the ${\bf S}^1$.  The result will be a sum of
logarithmic corrections
$$
	\log \left({M^{2} + p^2/R^{2} \over
\Lambda^{2}_{\rm UV}}
		\right)
$$
over $p$.
One can write this as the logarithm of a product and use the
product formula for $\sinh(x)$ to arrive at equation \pert; in
the process, one will have to discard an infinite piece
corresponding to the linear divergence in eq. \fdol.
In the limit $MR\to 0$ \pert\ becomes the expected
four-dimensional logarithmic correction;
$1/R$ takes the form of an ultraviolet cutoff
in this logarithm.

In five dimensions
the mass of a particle is the scalar component
of a vector multiplet. Upon the compactification on a
circle $\bf S^{1}$ the vector
multiplet gains an extra scalar - the Wilson loop around
$\bf S^{1}$
making the mass $M$ complex (the actual mass, entering the
Compton wavelength is the absolute value $\vert M \vert$). The
imaginary part
of $M$ is the vev of the Wilson loop. There are
five-dimensional
gauge transformations, which shift the value of the
Wilson loop by
the inverse radius $1\over{R}$. The formula \pert\ is
manifestly
invariant under such transformations:
$$
	M \to M + {{ip}\over{R}}, \quad p \in \IZ
$$
The same remark applies to  the coupling. In five
dimensions it is a
real scalar $g_{i}^{-2}$,
while upon compactification it gains the theta angle,
making up the holomorphic coupling
$$
	\tau_{i} = {{\theta_{i}}\over{2\pi}} + {{4\pi
		i}\over{g_{i}^{2}}}
$$
In compactifications with eight supersymmetries,
nonabelian gauge groups are generically broken
to their Cartan subalgebra via the Higgs mechanism,
due to the scalars in the same vector multiplet as the
gauge bosons.  The Weyl group should still be unbroken,
so in particular $M\to -M$ should be a symmetry, as it is in
eq. \pert.    

When we pass from M-theory on a Calabi-Yau manifold to type
IIA string theory on the
same Calabi-Yau, the apparent cutoff will be the
scale $g_{s}\ell_{s} = R_{11}$.
At weak string coupling the cutoff we might expect ($M_{s}$ or
maybe
$M_{p}^{(11)}$) is in fact at a lower scale than $1/R_{11}$.
Nonetheless, D-brane processes can involve momenta of
order $1/R_{11}$ even
for $R_{11} \ll \ell_{p}^{(11)}$ \shenkshort\dougshort,  
and so we
must include these
scales in our discussion.  We will meet this issue again when
discussing instanton corrections.

In general, there will be particles with various spins.
The effect
of the spin for a given particle
will be to add a factor in front of the
logarithm in \pert.
If the particle is in representation $r$ of the gauge
group $\CG$
and in representation $j$ of the massive little group,
then the
contribution is (see Chapter 16, section 6 of \peskinft\ for
a discussion.\foot{Here $d(j)$ is the dimension of
representation $j$.  For matrices $t^a$ representing the
Lie algebra
$\lieg$,
$C(r)$ is defined by ${\Tr} \left[t^a t^b\right] =
C(r)\delta^{ab}$.
Finally,
if $J^{\mu\nu}$ is a Lorentz generator in representation
$j$, then
${\Tr} \left[J^{\alpha\beta}J^{\mu\nu}\right] =
\left( g^{\alpha\mu}g^{\beta\nu} -
g^{\alpha\nu}g^{\beta\mu}\right)
C(j)$, and by making all of the indices spatial
one may derive an expression in terms of the representation
of the little group.  In five dimensions, we may write
$C(j) = {1\over 3} {\Tr} \left(\vec{J}_{1}^{2}
+ \vec{J}_{2}^{2}\right)$, where $\vec{J}_{1,2}$
are the generators of the two $\su(2)$ factors in the
little group \littlegroup.}
\eqn\spincoef{
	(-1)^{f}\left\{{d(j) \over 3} - 4 C(j)\right\}C(r)\ ,
}
where $f=0,1$ for
bosons and fermions respectively.

\subsec{Realizing five dimensional gauge theories in
IIA/M theory}

For $\su(2)$ gauge theories in
four dimensions an exact, nonperturbative (but implicit)
expression for the prepotential is known \SeWi\
(for a recent review and references see \lerchesw).  At
weak coupling, this
expression encodes the instanton corrections.
In five dimensions the gauge theory by itself is ill-defined,
since five dimensional gauge theories are nonrenormalizable.
To discuss them we must either work at one of the nontrivial
five-dimensional RG fixed points \seibfive, or embed
the theory in another theory which {\it is} well-defined
in the ultraviolet regime, such as string (and possibly
M-) theory.
The details of the string compactification will determine
the masses, charges, spins, and degeneracies
of BPS states which can flow in loops in the
five-dimensional theory.  For the rest of this section,
we will define the theory as type IIA
string theory compactified on  the local model described  
in the
introduction.  The answer we get will be different
for different local models (such as normal bundles to
$\IF_{1}$ of $\IF_{2}$ divisors of a compact
Calabi-Yau manifold \kkv), even though they may realize
the same four-dimensional gauge group.

The worldsheet instanton sum gives the complete answer
for the prepotential in $\CN=2$, $d=4$
compactifications of type IIA string theory.
Analyzing this prepotential gives some interesting results.
As a warm-up, take the limit $T_{b}\to \infty$ --
which is going to weak coupling from the gauge theory point
of view -- without necessarily scaling $T_{f}$ to zero.   
In this
limit, instantons wrapping around the base decouple.
Since $c_{0,m} = -2\delta_{m,1}$~\kkv,
the only contributions in the instanton
sum come from multiple covers of the fiber.
This sum is easy and we find that:
$$
	{2\pi i} \tau_{f} = -2 \log\left( 1 -
e^{-T_{f}}\right)\ .
$$
The authors of \kkv\ note that if we scale $T_{f}\to 0$ as
above, this
becomes the one-loop correction to the gauge
coupling in four dimensions,
as $T_{f}\propto M_{W}$.
However, before taking this scaling limit
we can see that up to a piece linear in
$T_{f}$, this correction is:
\eqn\noinstcorr{
	{2\pi i}\tau_{f} = -2 \log\sinh^{2}\left(
	{T_{f}\over 2}\right) ,
}
which is the result of a loop integral on 
$\IR^{4}\times {\bf S}^{1}$, 
provided $T_{f} = 2\pi M_{W} R$.
The missing piece is the one-loop correction in
five uncompactified dimensions.  In M-theory
compactifications on Calabi-Yau
threefolds, this one-loop correction
shows up in the classical prepotential 
\fivedcube\seibmor\dkvdp\seibmorint (close to
the enhanced gauge symmetry point and at weak coupling).
Our interpretation of eq.
\noinstcorr\ makes a bit more sense if we note that
$$
	M_{W} = {1\over{2\pi}} {T_{f} \over \ell_{s}g_{s}}\ ,
$$
as the $W$-boson is a wrapped D2-brane.  
We know that type IIA
string theory is dual to
M-theory on an additional ${\bf S}^{1}$, and the radius
of this ${\bf S}^{1}$ is $R_{11}=g_{s}\ell_{s}$.
So we get the expected expression,
$\log\sinh^{2} \left( {\pi} M_{W}R_{11} \right)$.
The relation between the five- and four-dimensional
couplings is:
$$
{1\over{g_{4}^{2}}}  \sim {{2\pi R}\over{g_{5}^{2}}}
$$
which is why \noinstcorr\ is missing the the factor of  
$1/2\pi R$ that appears in \pert.

The fact that classical string theory reproduces
loop effects of D-branes is no
surprise~\conif.  In this example the relation
between these two calculations is easy to see.
The one loop
answer in the gauge theory comes from M-theory 2-branes
wrapped around the fiber;
the terms $e^{-2\pi kM_{W}R_{11}}$ come from worldline
instantons wrapping $k$
times around the additional ${\bf S}^{1}$ of the
M-theory compactification.  From the
string theory point of view
these are strings wrapping $k$ times around the
fiber and once around
the ${\bf S}^{1}$.
But these configurations are the same \beckstrom:
multiple covers are really
multiple instantons.  The same is true
for multiple windings of worldline
instantons.  Thus, M-theory provides a
geometrical picture of how string effects resum to
D-brane effects.

At this point the reader might
object that the prepotential should be independent of  
$R_{11}$, since
the type IIA dilaton lives in a
hypermultiplet.  The prepotential is in fact independent
of $g_{s}$ when we express the prepotential in terms of the
appropriate
coordinates on the vector multiplet moduli space.
The radius enters when we try to identify
these coordinates with masses and energy scales that
we might measure in physics experiments.
(This identification also leads to the emergence
of $R_{11}$ as the ultraviolet cutoff near the
conifold point of type IIB compactifications on
Calabi-Yau threefolds \shenkshort.)

Now let us look at the full worldsheet instanton sum for
our noncompact Calabi-Yau example.  Returning to
eq. \fnoughtinst\ and summing over $k$, we find that
up to terms linear in $T_{b,f}$:
\eqn\fullprepot{
	2\pi i \tau_{f} = \sum_{n,m} c_{n,m} m^{2}\log\sinh^{2}
\left( {nT_{b}+m T_{f} \over 2}\right)
}
This looks like a sum of one-loop integrals.  We can give
this sum the same interpretation we gave
the weak-coupling limit in the previous paragraphs.
We are missing an important piece, though, namely
the linear terms promoting
$\log(1 - e^{-nT_{b}-mT_{f}})$ to the summands in eq.
\fullprepot.  At present we do not understand their origin;
they will become important when the entire $\IF_{0}$ divisor
shrinks to a point.

If we think of the type IIA compactification
as an M-theory compactification on an additional
${\bf  S}^{1}$, we
know that M2-branes
wrapped around holomorphic curves in the homology class
$$
	n [\beta_{b}] + m [\beta_{f}]\in  
H_{2}(\CN(\IF_{0}),{\IZ})\ ,
$$
give rise to BPS states.  (Here $[\beta_{b,f}]$ are the
homology classes of the base and fiber of the Hirzebruch
surface.)
We label these states by their bidegree $(m,n)$. Their
central charges are:
$$
Z_{m,n} \propto (nT_{b}+mT_{f})/g_{s}\ell_{s}.
$$
Their mass is therefore
$M_{(n,m)}= \vert  nT_{b}+mT_{f} \vert /g_{s}\ell_{s}$,
so that the
arguments of the $\log\sinh$ terms
look like $M_{(n,m)}R_{11}$ as expected.
The $m^{2}$ part of \fullprepot\ is simply the charge of the
2-brane under $\CU(1)_{f}$.

Furthermore, we claim that
the branes will carry instanton number $n$.  The simplest
way to see
this is to note that in M-theory compactifications on
Calabi-Yau manifolds, there are Chern-Simons couplings of  
the form \cadavid\papatown\gunasiertown
\eqn\chsim{
	\CS_{{\rm Chern-Simons}} \propto
		\int_{\IR^{5}} C_{ijk} A^{i}\wedge F^{j}
		\wedge
		F^{k}\ .
}
The models we are interested in are fibrations of
ALE spaces over the base $\IP^{1}$.  By analogy with the
compact Calabi-Yau case \asplouis, we expect that the
intersection numbers
will include terms of the form
$$
	C_{bij} = \eta_{ij}\ ,
$$
plus terms independent of $[\beta_{b}]$.
Here $\eta$ is the intersection matrix for
divisors dual to cycles of the ALE fiber which are
invariant under the monodromy group
of the fibration, and has signature $(1,-1,\ldots,-1)$.
There are of course subtleties here since the
ALE space is noncompact.  In particular, there will
be a vector multiplet arising from the Kaluza-Klein mode in
the noncompact direction.  If we have reached
this noncompact example by making a K\"{a}hler class large,
it will be the scalar corresponding to this class; call it
$t_{n.c.}$.  Thus we
still expect the scalar of this multiplet to show up in  
the cubic
terms of the prepotential.  Since we are looking
at an example with a low-rank gauge group, we will assume  
that this
multiplet arises from the scaling limit of a compact  
Calabi-Yau.
Our $\IF_{0}$ example is contained in the degree 24  
hypersurface in
$\IP_{1,1,2,8,12}$ as the base of an elliptic fibration, and
may be reached by making the elliptic fiber large.  The
divisor dual to the elliptic fiber
will be the base $\IF_{0}$.  The divisor dual to the
base $\IP^{1}$ will be the remnant of generic the K3 fiber,
which is the
bundle $\CO(-2)$ over the fiber $\IP^{1}$; similarly, the dual
to the fiber will be the line bundle $\CO(-2)$ over
the base.\foot{We would like to thank P. Aspinwall
for suggesting these identifications.} 

Close to the enhanced gauge symmetry point, and at
weak coupling, the authors of \seibmor\dkvdp\ have
shown that there is a term of the form
$$
	\int_{\IR^{5}} A^{b}\wedge F^{f}\wedge F^{f}
$$
in the Lagrangian (where $A^{b}$ is in the vector multiplet
with $t_{b}$.) The topological current
\eqn\instcurr{
	j = {\star} F^{f}\wedge F^{f}
}
is globally conserved (\cf\ \seibfive);  this charge is just
the instanton number.  Eq. \chsim\ couples $A^{b}$
to this charge; the equations of motion
then state that field configurations with charge $n$
under this $\CU(1)$ also have instanton number $n$
with respect to $\CU(1)_{f}$.  Note that a related  argument
for the type IIA picture
was made in \kkv  to show that worldsheet instantons
wrapping around the base carried instanton number
with respect to the fiber $\IP^{1}$.  This
argument began with the term $B\wedge F \wedge F$ that
appears in IIA compactifications on $\bf K3$ \vafwitol.
Both this term and the term in eq. \chsim\ arise from
the same term in eleven dimensions.

Finally, we would like to interpret the coefficients
$c_{n,m}$ in terms of the five-dimensional gauge theory.
The latter suggests that this number is more than just
the numbers of curves of bi-degree $(n,m)$.  Since the wrapped
two-branes will generically lead to the states with  spin,
there will be contributions to $c_{n,m}$
of the form \spincoef.
We may follow arguments due to Witten \phasemf\ in
finding this
contribution.  For $\CN=1$ theories in five dimensions,
the supercharges transform as $2(1/2,0)\oplus 2(0,1/2)$
under the little group \littlegroup.  Since the
two-branes are wrapped around holomorphic curves they give
rise to BPS states breaking half of the supersymmetries
\beckstrom; let them break the supercharges transforming
nontrivially under $\su(2)_{2}$. The broken supercharges give
rise to four fermion zero-modes; upon quantizing
them we can write them as two pairs of fermionic creation
and annihilation operators.  Allowing these operators to
act on the obvious  vacuum state. we
get two bosonic states and two fermionic states which
must transform as $2(0,0)\oplus(0,1/2)$ -- \ie\ they form a
``half-hypermultiplet.'' In addition, the two-brane  
configurations
will have a moduli space $\CM_{(n,m)}$,
and the fermionic partners (under the unbroken supersymmetry)
of the moduli space coordinates give rise in the usual
way to forms on $\CM_{(m,n)}$ (for a recent review, see
\newton).  The supersymmetry
operators act as differentials
on this space, and BPS states are harmonic forms on the
moduli space. As $\CM_{(m,n)}$ is K\"{a}hler,
the Dolbeaux cohomology of this moduli space has a
natural $\su(2)$ action \lefshetz,
provided the space has a proper compactification (which is
the case when this moduli space is the
moduli space of worldsheet instantons);
we can write the action such that forms in $H^{p,q}$ have
$J^{(3)}$ eigenvalue
$$
	\half (p+q- {\rm dim} _{\IC} {\CM}_{n,m})\ .
$$
$J^{+}$ acts on forms via the wedge product with the
K\"{a}hler form $\omega$ of $\CM_{(m,n)}$
and $J^{-}$ is its adjoint with respect to the Hodge pairing.
In fact this $\su(2)$ is just $\su(2)_{1}$.\foot{Had
we studied the BPS states in the type IIA theory on $\bf K3$, the
corresponding moduli space of curves would be hyper-K\"ahler
and the little group $\CS\CO(5)$ would
act on its cohomology in accordance with \roslhyper\verbit}
Forms of odd degree are spacetime
fermions and forms of even degree are spacetime bosons.
We can organize the cohomology into $\su(2)$ multiplets
which will transform as
$(k/2,0)$ under \littlegroup.  The complete spectrum
comes from tensoring
the representations for a given $k$ with the
half-hypermultiplet
coming from the broken supersymmetries leading to
states in the representation $2(k/2,0)\oplus(k/2,1/2)$.
For this representation the contribution from \spincoef\
is just
$$
	2(-1)^{k/2}(2 {k\over 2} + 1)\ .
$$
$(-1)^{k/2}$ counts the ${\IZ}_{2}$ grading of the degree of
the forms on $\CM_{m,n}$
contributing to this $\su(2)$ multiplet.
Thus, summing over all of the BPS states at this degree
(as we only have to sum over BPS states \hm),
we find that up to numerical factors independent of $(n,m)$,
\eqn\braneeuler{
	c_{n,m} = \chi (\CM_{n,m})\ .
}
This is the result we expect from string theory \AandB,
but with an new interpretation; the Gromov-Witten
invariants are the contributions of wrapped BPS
two-branes to the one loop
beta-function of the five dimensional theory.
\foot{Earlier work using mirror symmetry to
count the degeneracy of wrapped 2-branes
can be found in \kmvbps.  (See also
subsequent work in \lmwnoncrit\mnwnoncrit.) In this
problem we find that the relevant Euler character
is not counting the degeneracy of BPS states, but rather
the weighted sum that appears in front of the one-loop 
$\beta$-function.}
However, just as in the discussion of the weak-coupling
limit of \fnoughtinst, the relation of M-theory to
type IIA string theory means that these should be
equivalent ways of describing the expansion of the
prepotential in $e^{-T_{b}}$, $e^{-T_{f}}$.

\subsec{Scaling limits of string- and M-theory
compactifications}

We have already argued that the K\"{a}hler
class $T_{b}$ of the base of the fibration
is the gauge theory coupling \asplouis:
$$
T_{b} = {1 \over g^{2}}\ .
$$
(The gauge theory coupling $g$ is distinct from the type
IIA string coupling $g_{s}$.) The charged gauge bosons
arise from D2-branes wrapping around the appropriate  
2-cycle in
the $\bf K3$ fiber \bsvegs; their masses are
\eqn\wmassmod{
	{{T_{f}^{i}}\over{\ell_{s} g_{s}}} = M_{W}^{i}\ .
}
Written in terms of physical parameters, eq. \fullprepot\
becomes
\eqn\physicalcoupl{
	\tau_{f} = -i \sum_{n,m} c_{n,m} m^{2}\log \sinh
		\left( {M_{(n,m)}R_{11}\over 2}\right)\ ,
}
where $M_{(n,m)}$ is the mass of the two-brane wrapped around
the cycle $n[{\rm base}] + m[{\rm fiber}]$.
$M_{(n,m)}$ and $R_{11}$ will be different numbers
depending on whether we measure them with
the type IIA metric or $11d$ supergravity metric, but
their product is a ratio of length scales and
this will be invariant under metric rescalings.

In four dimensions, dimensional transmutation relates
the gauge coupling to the QCD scale 
$\Lambda_{{\rm QCD}}$.  In
order to make this identification we need to
know at what ultraviolet scale the gauge coupling
is defined.  Of course, string theory is a finite
theory, so that the gauge coupling is what it is.
However, when discussing the low energy physics we
integrate out physics above the low energy scale we
are interested in, and if we wish to
discuss renormalization group invariants (and thus
appeal to dimensional transmutation) we
need to know up to what scale we are integrating
out virtual processes.  Once we know $\Lambda_{{\rm UV}}$,
the fact that
$$
	\Lambda^{4}e^{- {1 \over g^{2}(\Lambda)}}
$$
is an exact renormalization group invariant for $d=4$
supersymmetric theories means that we can write
$$
	\Lambda_{{\rm QCD}}^{4} =
	\Lambda^{4}_{{\rm UV}} 
	e^{- {1 \over g^{2}
		(\Lambda_{{\rm UV}})}}\ .
$$
Naively $\Lambda_{{\rm UV}}=m_{s}$,
but given the results of \shenkshort\dougshort\
this is a questionable assumption.
In particular, we have seen that
our analysis of the weak-coupling limit naturally
identifies $1/R_{11}$ as the UV scale.
We will now analyze the instanton sum to show that
the gauge coupling is also naturally defined at this scale.

In order to extract the four-dimensional instanton sum,
\kkv\ force $T_{f}^{4}$, $e^{-T_{b}}$ to become small at
the same rate by making them both proportional to a
small dimensionless parameter $\epsilon^{4}$.
For the example we have been studying, the full instanton sum
\fnoughtinst\ becomes
\eqn\fnoughtlim{
	\CF = \sum_{n} c_{n} \left( {e^{-T_{b}} \over
		T_{f}^{4}} \right)^{n}
}
to leading order in $\epsilon$.
This should be the familiar spacetime instanton expansion,
once we have properly identified the Calabi-Yau moduli with
physical parameters in the gauge theory.
Substituting in \wmassmod, eq. \fnoughtlim becomes
$$
	\CF = \sum_{n} c_{n} \left( {e^{-T_{b}}
	\over \ell_{s}^{4} g_{s}^{4} M_{W}^{4}}\right)^{n}\ .
$$
We can rewrite this as
$$
	\CF = \sum_{n} c_{n} \left(
		{\Lambda_{{\rm QCD}} \over M_{W}} \right)^{4n}
$$
if we identify the UV scale with
$$
	\Lambda_{{\rm UV}} =
		{1\over g_{s}\ell_{s}} = {1\over R_{11}}\ .
$$
Then the QCD scale $\Lambda$ is:
$$
	\Lambda_{{\rm QCD}}^{4} = 
	{m_{s}^{4} \over g_{s}^{4}} e^{-T_{b}}\ .
$$
The ultraviolet scale shown here is exactly the ultraviolet
scale found in \shenkshort.  In fact, our arguments are
simply the weak coupling version of the arguments there.
The fact that the gauge coupling is naturally defined up to
this scale is no surprise given the results of \dougshort;
there are processes with momentum exchange up to this scale,
so we need to start integrating
out virtual processes from this scale down to the low  
energy scale
in order to recover the (Wilsonian) low-energy effective  
action.

Given the above identifications, we can see directly that the
scaling limit in \kkv\ is the four dimensional limit
of a five dimensional theory.  We have just shown that
\eqn\rscaling{
	\eqalign{
		T_{f} &= M_{W} R_{11} \cr
		e^{-T_{b}} &= \left( \Lambda_{{\rm QCD}} R_{11}
	\right)^{4}\ .
	}
}
Given this identification of the moduli with physical
gauge theory parameters,
the scaling $R_{11} \sim \epsilon$ then reproduces the
scaling limit of \kkv.

\newsec{Local mirror symmetry and relativistic integrable
systems}

To date the most powerful method for calculating
worldsheet instanton
corrections is mirror symmetry.
The cases for which finding the mirror  manifold
is more than an art are intersections of hypersurfaces in
toric varieties.  In these cases, the mirror is
straightforwardly extracted
from the toric data describing the manifold, as described
by Batyrev
\batduke\batdual\bathyperg.
For noncompact
Calabi-Yau toric varieties this story is less clear.
Batyrev's presentations depends on relating monomial
deformations of the defining polynomial of a hypersurface
to K\"{a}hler deformations of the mirror, but for
our example there are no such monomial deformations.
The authors of \kkv\ have argued that the relevant
mirror geometry for noncompact Calabi-Yau toric varieties is
a Riemann surface.  In cases that the
singularity of the noncompact model can be imbedded in a
compact toric variety, this surface sits naturally inside
the compact Calabi-Yau.  However, a local model can
have a singularity corresponding to a gauge group
of arbitrary rank and there may not be
a compact Calabi-Yau containing such a singularity,
so that it is of some interest to find the answer directly
from the noncompact manifold.
In fact, Batyrev's conjectures really
provide the Picard-Fuchs equations (and thus
the relevant hypergeometric functions) for
the periods of the original Calabi-Yau surface.
If we simply extract these hypergeometric functions
we will find that they have a natural integral representation
as periods of a Riemann surface.  We will find, however,
that blindly applying these techniques causes us to miss an
important piece of the physics, namely the dependence of the
prepotential on the Kaluza-Klein modes along the noncompact
direction.
We will find that including this scalar is necessary if  
the periods
we
derive are to be related to a prepotential.

\subsec{Extracting the local mirror}

Givental \given\giventwo\ claims to have proven
that the genus zero worldsheet instanton
corrections for toric varieties and
hypersurfaces in toric varieties can be described by  
appropriate
hypergeometric functions as described in
\batduke\bathyperg.
His method involves localization formulae
for the equivariant Gromov-Witten invariants.  A full
discussion of toric varieties, localization, and  
Givental's proof
will appear in \wepromise; for now we will simply use
his presentation of the relevant hypergeometric functions.

For the purposes of this discussion let us concentrate on the
local model $\CN(\IF_{0})$.  This can be described as a  
noncompact
Calabi-Yau toric variety with the charge vectors:
\eqn\fnotcharge{
	\left\lgroup\matrix{ Q^{(1)}_{k=1 \cdots s}
		\cr Q^{(2)}_{k=1 \cdots s} }\right\rgroup =
\left\lgroup\matrix{1&1&0&0&-2\cr 0&0&1&1&-2}\right\rgroup\ ,
}
describing a $\CU(1)^{2}$ action on $\IC^{5}$.
(Our notation follows that of \morple, which also discusses
noncompact
toric varieties.  Other nice discussions of toric geometry for
string theorists, with ample references, are contained in
\agmbig\ and \hosonorev.)
The generating function for the prepotential and the periods
for the K\"{a}hler moduli is (the coordinates $t_{1,2}$  
are different
from $T_{1,2}$!):
\eqn\gvntl{I(\{t\}; x_{1}, x_{2}) = \sum_{a_{1}, a_{2} \geq 0}
	e^{- t_{1} (x_{1} + ta_{1}) - t_{2} (x_{2} + ta_{2})}
	{{\prod_{\ell=0}^{2a_{1}+2a_{2}-1} (2{ x_{1} +  
2x_{2} + \ell t )
	}\over{\prod_{m=1}^{a_{1}} (x_{1} + m t )^{2}
	\prod_{n=1}^{a_{2}}
	(x_{2} + n t )^{2}}}}
}
The periods are found by expanding in $x_{1,2}$ mod
$x_{1}^{2}, x_{2}^{2}$.  Givental
\given\giventwo\ has shown that this generating
function has the form
\eqn\gvntlspec{
	I = e^{-(T_{0} + T_{1}x_{1} + T_{2}x_{2})}
		\left( 1 + o(1/t) \right)\ ,
}
where $T$ are the ``nonlinear $\sigma$-model coordinates''
\agmbig\morple.  We can read the change of coordinates off
of the terms in \gvntl\gvntlspec\ linear in $x/t$, and for
hypersurfaces in toric varieties this specifies the  
mirror map. Here $T_{0}$ is the fundamental period
which in our discussion vanishes.

We can also derive
the Picard-Fuchs equations from the generating function
in \gvntl.  For the case at hand the equations are as  
given in \kkv:
\eqn\fnoughtpf{
	\eqalign{
	\CD_{1} &= \partial_{t_{1}}^{2} - q_{1}\left(
		2\partial_{t_{1}} + 2\partial_{t_{2}} +  
t\right)
		\left(2\partial_{t_{1}} +  
2\partial_{t_{2}}\right)
			\cr
	\CD_{2} &= \partial_{t_{2}}^{2} - q_{2}\left(
		2\partial_{t_{1}} + 2\partial_{t_{2}} +  
t\right)
		\left(2\partial_{t_{1}} +
			2\partial_{t_{2}}\right)\ .
	}
}
These equations have a constant solution.  The operators  
$\CD_{1,2}$
annihilate \gvntl\  mod $x_{1}^{2}, x_{2}^{2}$.
In fact, the recursion relations for the coefficients of
$e^{-(d+x)t}$ in
eq. \gvntl\ leads to the equations
$$
	\CD_{1,2} I = x_{1,2}^{2}
G(t_{1},t_{2})\ .
$$
Thus, only the coefficients of  
$(1,x_{1},x_{2},x_{1}x_{2})$ are annihilated by 
$\CD_{1,2}$.  Indeed, as $x_{1},x_{2}$  
are the basis
for $H^{2}(\IF_{0})$ they satisfy the relations  
$x_{1,2}^{2}=0$ (\cf\ \morple).

$T$ is easily seen to be
\eqn\linlog{
 	T_{1,2} = t_{1,2} - 2 f(t_{1}, t_{2})
}
where
\eqn\linlogseries{
	f(t) = \sum_{a_{1}, a_{2} \geq 0}
	{{(2a_{1}+2a_{2}-1)!}\over{(a_{1}!)^{2} (a_{2}!)^{2}}}
	q_{1}^{a_{1}} q_{2}^{a_{2}}\ .
}
The last non-trivial solution of $\CD_{1,2}=0$ is the  
coefficient of
\gvntl\ in front of $x_{1}x_{2}$:
\eqn\lstprd{T^{D} = t_{1}t_{2} - 2 (t_{1} +t_{2}) f(t) -  
4(g_{1} +
g_{2}) = T_{1}T_{2} + 4(f^{2}- g_{1} - g_{2})}
where:
\eqn\tdual{
	\eqalign{
		& g_{1}(t) = \sum_{a_{1}, a_{2} \geq 0}
	 {{(2a_{1}+2a_{2}-1)!}\over{(a_{1}!)^{2}
		(a_{2}!)^{2}}}
	\left( \sum_{m=a_{1}+1}^{2a_{1}+2a_{2}-1}  
{1\over{m}}  \right)
q_{1}^{a_{1}}q_{2}^{a_{2}}
		\cr
	& g_{2}(t) = \sum_{a_{1}, a_{2}\geq 0}
		{{(2a_{1}+2a_{2}-1)!}
		\over{(a_{1}!)^{2} (a_{2}!)^{2}}}
	\left(	\sum_{m=a_{2}+1}^{2a_{1}+2a_{2}-1}
			{1\over{m}}  \right)  
q_{1}^{a_{1}}q_{2}^{a_{2}}\cr
	}
}
The four periods $(1,T_{1},T_{2}, T^{D})$ are all of the  
solutions
to the Picard-Fuchs equations \fnoughtpf.  At this point there
appears to
be a problem, as we expect a term cubic in $T$ to serve as
the prepotential, and we would expect two dual periods  
quadratic
in $T$.  The resolution has been discussed already,  
namely that
these equations do not take into account the additional scalar
$T_{3}$ coming from the Kaluza-Klein mode in the noncompact
direction.
This will not appear in the instanton sum but it will  
appear in the
prepotential.  One can convince oneself of this by  
imbedding the
local model into a known compact Calabi-Yau.  For example,
the local model of $\IF_{2}$ discussed in \kkv\ can be  
imbedded
in the degree 24 hypersurfaces in $\IP^{4}_{1,1,2,8,12}$,
at the point where this manifold can be described as an  
elliptic
fibration
of $\IF_{2}$.  (This model has been discussed in \hosonoone.)
There are three
K\"{a}hler classes:  $T_{f,b}$ measures the sizes of the  
fiber and
base of the $\IF_{2}$, and $T_{k}$ measures the size of the 
{\bf K3}
which is the elliptic fibration of  the fiber of $\IF_{2}$.
(The corresponding algebraic coordinates
$t_{k},t_{b},t_{f}$ are called $\log x,\log y,\log z$ in  
\hosonoone.)

In the limit $T_{k}\to \infty$, the instanton corrections  
to the
prepotential
will not depend on $T_{k}$.  By additionally demanding that
the periods do not depend on $T_{k}$, we find that they will
satisfy the Picard-Fuchs equations for the local model  
described in \kkv;
and indeed there are only four solutions, 
with the terms polynomial  in $T$
proportional to $(1,T_{f},T_{b},T_{f}^{2}-T_{f}T_{b})$.   
The last solution (the solution with a term quadratic in $T$) is:
$$
	\left({\partial \over \partial T_{k}} - 2  
{\partial \over \partial
T_{b}}\right)
		\CF(\{T\})\ .
$$
We can see in this example that by blindly extracting the
Picard-Fuchs
equations for noncompact toric varieties, we will miss an  
important
piece of the physics.  We hope to return to this issue in  
the future
\wepromise.

Now let us examine $T_{1}$, represented as the infinite  
series:
\eqn\hypr{
	T_{1} = t_{1} - 2\sum_{a_{1}, a_{2} \geq 0}  
D_{a_{1}, a_{2}}
	q_{1}^{a_{1}} q_{2}^{a_{2}}
}
Recall that $D_{a_{1}, a_{2}} =
{{(2a_{1}+2a_{2}-1)!}\over{(a_{1}!)^{2} (a_{2}!)^{2}}}$.
It is quite easy to relate this series to a period
of the (non-compact) elliptic curve \nikitafive :
\eqn\crve{
	F(w_{1}, w_{2}) \equiv \sqrt{q_{1}} \left( w_{1}  
+ {1\over{w_{1}}}
\right) +
	\sqrt{q_{2}} \left( w_{2} + {1\over{w_{2}}}  
\right) = 1
}
sitting in the $(w_{1}, w_{2})$ plane. Start with the  
identity:
$$
	\sum_{\matrix{& a,b,c,d \geq 0\cr
	& a + b +c +d > 0\cr}} X^a Y^b Z^c Q^d
	{{(a+b+c+d-1)!}\over{a!b!c!d!}} = - {\rm log} (1  
- X - Y - Z - Q)
$$
Now put $X= \sqrt{q_{1}} w_{1}$, $Y = \sqrt{q_{1}}  
w_{1}^{-1}$,
$Z = \sqrt{q_{2}} w_{2}$, $Q = \sqrt{q_{2}} w_{2}^{-1}$.  
Let us
suppose that $\vert q_{1,2} \vert << 1$. Take the contour  
integral
$$
	{{dw_{1}}\over{w_{1}}}  {{dw_{2}}\over{w_{2}}}
$$
around $w_{1} = w_{2} =0$ along a contour $\vert w_{1,2}  
\vert =
\epsilon_{1,2}$,
such that $ \epsilon_{1,2}, {{\vert q_{1,2}
\vert}\over{\epsilon_{1,2}}} <<1$. The contour
integration
enforces the constraint $a=b$, $c=d$. Thus,
\eqn\prd{f(t) = - {1\over{(2\pi i)^{2}}} \oint
{{dw_{1}}\over{w_{1}}}
\wedge {{dw_{2}}\over{w_{2}}}
\, {\rm log} \left(1 - F(w_{1}, w_{2}) \right)}
which makes $T_{1}$ a  period of the restriction of the
differential
\eqn\dfnr{\lambda = -2 {1\over{2\pi i}} {\rm log} (w_{1} )
{{dw_{2}}\over{w_{2}}}  }
onto the curve $F = 1$.
This curve is exactly the curve derived for $\IF_{0}$
in \kkv.

\subsec{The four-dimensional limit}

As $R\to 0$ our Riemann surface representation of
the periods of the prepotential should become the
Seiberg-Witten representation \SeWi.
Start by writing the parameters in \hypr\ and \crve as:
\eqn\sclngs{
	\eqalign{
	& q_{1} = {1\over{4}} + R^{2} u +
		\ldots\quad
		q_{2} = R^{4} \Lambda^{4}\cr
	& w_{1} = 1- 2i R p + \ldots  \quad w_{2} =
		R^2 \chi + \ldots\ ,\cr
	}
}
and take the limit $R \to 0$.
We anticipate that near $q_{1} \sim {1\over{4}}$ the
series \hypr\
is about
to diverge, and the most important contribution comes
from the terms
with very large $a_{1}$.
Asymptotically:
\eqn\asmpt{D_{a_{1}, a_{2}} \approx {{4^{a_{1}+a_{2}}  
a_{1}^{2a_{2} -
{3\over{2}}}}\over{2\sqrt{\pi} (a_{2}!)^{2}}}, \quad a_{1} \to
\infty}
(compare to the prediction $c_{a_{1}, a_{2}} \propto
a_{1}^{4a_{2}-3}$ of \kkv).
Thus $f(t)$ behaves as:
\eqn\ashypr{
	\eqalign{
	f(t) &\sim \sum_{a}
	{{(4R^{4}\Lambda^{4})^{a} }\over{2\sqrt{\pi}  
	(a!)^{2}}}  
	\sum_{b}
	b^{2a - {3\over{2}}} ( 1 + 4R^{2}u)^{b} \cr
	& \sim  i R
	\sqrt{{u}\over{\pi}} \sum_{n = 0}^{\infty}
		{{(2n-{3\over 2})!}\over{(n!)^{2}}} \left(
		{{\Lambda^{2}}\over{2u}} \right)^{2n}\cr}}
which is one of the periods of the differential
\eqn\swdfr{\lambda_{SW} = -2{1\over{{\pi} }} R p
{{d\chi}\over{\chi}}
}
The curve \crve\ goes in this limit to  the  
Seiberg-Witten curve:
\eqn\swcrve{
	- p^{2} + u + {\half} \left( \chi +
	{{\Lambda^{4}}\over{\chi}} \right) = 0
}

\subsec{Relations to integrable systems,
	and generalization to $\su(n)$}

The Seiberg-Witten curve \swcrve\ 
above is the spectral curve
for the periodic $A_{1}$ Toda lattice.  In general it is now
known that the periods of the
prepotential for four-dimensional $\CN=2$ gauge theories
with gauge group $\CG$
are periods of the Riemann surface described by the
spectral curve for the periodic Toda lattice living on
the Dynkin diagram dual to the diagram for $G$
\emilnickone\gmmm.  The differential
one integrates in just the action differential $p{\rm d}q$
of the integrable system.  One of the present authors
\nikitafive\ conjectured that one may
recover the prepotential
for the five dimensional $\CN=1$ gauge theory on
$\IR^{4}\times {\bf S}^{1}$ from the periods of the
spectral curve for the relativistic Toda lattice
described in \ruijone.  For $\su(2)$ the curve
and differential are exactly as presented in eqs.
\crve\dfnr.  It is clear that the curve and periods
will in general be highly dependent on the details
of our realization of gauge symmetry in the
Calabi-Yau compactification.  For example, we
may realize $\su(2)$ via noncompact Calabi-Yaus
containing any of $\IF_{0,1,2}$ as divisors.
When one extracts the four-dimensional answer via a
scaling, all of these compactifications lead to
the Seiberg-Witten story.  Away from this
scaling limit, however, one will find very different
curves.  This is hardly a surprise; since five-dimensional
theories are nonrenormalizable the prepotential will be
very dependent on the microscopic details of the theory.
On $\IR^{4}\times {\bf S}^{1}$ the effects which are
nonperturbative
from
the four-dimensional point of view depend on the details
of the
particle spectrum in five dimensions which in M-theory
compactifications will depend on the details of the
compactification manifold.

The $\su(n)$ generalization described in \nikitafive\ can
also be realized via the same techniques described above.
The relevant Calabi-Yau toric variety can be described
via the charge vectors
\eqn\suncharges{
	\left\lgroup\matrix{ Q^{(1)}_{k=0 \cdots n+2} \cr
		Q^{(2)}_{k=0\cdots n+2} \cr
		\vdots\cr
		Q^{(n)}_{k=0 \cdots n+2}}
	\right\rgroup =
	\left\lgroup\matrix{1&1&-1&0&0&0&0
				&\ldots&0&0&-1 \cr
			0&0&1&-2&1&0&0&
				\ldots&0&0&0\cr
			\vdots\cr 0&0&0&0&0&0&0&
			\ldots&1&-2&1}\right\rgroup\ .
}
acting on
$$
	z_{0},z_{1},y_{0},\ldots,y_{n}\ .
$$
This is an $A_{n-1}$ fibration over $\IP^{1}$.
Using the methods of the previous section,
we find that the periods of the prepotential
are described by the Riemann surface described in \nikitafive,
\eqn\sunsurf{
	b(z + {1\over z}) +
y^{-n/2}\sum_{k=0}^{n}\alpha_{k}y^{k} = 0\ ;
}
to get the periods we integrate
$$
	\lambda = {1\over 2\pi i} \log(z) {{\rm d}y \over y}
$$
over the curve \sunsurf.

\newsec{Comparison of the instanton sum in the heterotic dual}

There is ample evidence that the heterotic string
compactified on $\bf K3\times {\bf T}^{2}$ is dual to the
type IIA string compactified on a $\bf K3$-fibered Calabi-Yau
threefold \fhsv\kachruvafa\asplouis\  (for a thorough
review and references see the lectures \asplect.)
In general, no explicit map between the worldsheet instanton
sum of a type IIA compactification and the spacetime
instanton
sum on the heterotic side is known for $\CN=2$
compactifications.
For $\CN=4$ compactifications this identification has
been made
in \hmrtwo, where the authors provided some evidence that
genus one worldsheet instantons on the ${\bf T}^{2}$ factor of
the type IIA compactification (on $\bf K3\times {\bf  T}^{2}$)
were mapped to
the heterotic five-brane wrapped around ${\bf T}^{6}$.
The authors
of \hmrtwo\ then made the argument that the
$\CN=2$ case could be described as follows.  If we begin
with the $\so(32)$ heterotic string compactified on 
${\bf  T}^{4}$,
we can find the dual type IIA string compactified on $\bf K3$
by wrapping the ``small instanton'' five-brane around the
torus \smallinst.  The fivebrane has an $\su(2)$ gauge  
group on
its worldvolume and the Wilson lines around the ${\bf T}^{4}$
have the moduli space ${\bf T}^{4}/{\IZ}_{2}$, a special
$\bf K3$ but a $\bf K3$ nonetheless.  One may understand
heterotic-type IIA duality
in $d=4$, $\CN=2$ compactifications by adiabatically fibering
the $\bf K3$ of the type IIA theory and a ${\bf T}^{2}$
factor of the
heterotic theory over a $\IP^{1}$ \vafwit.  A worldsheet
instanton of the type IIA string theory wrapped around
the base of the resulting $\bf K3$ fibration will then look
on the heterotic side like a heterotic fivebrane wrapped
around ${\bf K3}\times {\bf T}^{2}$.

However, there is more to the worldsheet instanton sum in  
these
Calabi-Yau models than the instantons wrapping around the
$\IP^{1}$ base of the $\bf K3$ fibration.
If the type IIA compactification has
enhanced gauge symmetry corresponding to perturbative
enhanced gauge symmetry in
the heterotic dual, the gauge symmetry will arise from
$\IP^{1}$s in the generic $\bf K3$ fiber (more precisely, in
a monodromy-invariant subgroup of the Picard lattice
of the fiber).  As we have discussed
in the previous section, the worldsheet instantons
wrapping these $\IP^{1}$s can also be described as worldvolume
instantons of
M2-branes wrapped around  $\bf S^{1} \times \IP^{1}$.
The $W$ bosons on the heterotic side correspond to heterotic
strings and so we might expect the full instanton sum on the
heterotic side to be
built up from $5$-brane and string instantons; the fivebrane
provides a ``core'' for the instanton solution, carrying
the instanton number, and the full solution is built up by
a superposition of heterotic worldsheet instantons.

To get better intuition for this it is useful, following the
analysis of the previous section, to lift the discussion to
five dimensions.  It has been conjectured
that M-theory compactified on a Calabi-Yau manifold is
dual to the heterotic string compactified on
${\bf K3} \times {\bf S}^{1}$ \antfertay.  We will see
how this can be deduced
from the four-dimensional duality in an example below.
On the M-theory side we know that the instanton sum we get by
compactifying the theory on an additional circle comes
from two-branes wrapped around cycles of the Calabi-Yau.
If we can match these particles to particles on the heterotic
side, we can then match the instanton sum in four dimensions
once we compactify on an additional circle, as in both
theories the instantons will arise from worldline instantons
of the five-dimensional particles.  These particles will
be bound states of heterotic strings and fivebranes.
Computing their degeneracy and spin seems
quite difficult.  On the other hand, if we
are wiling to accept an additional duality conjecture, namely
heterotic/type $I$ duality, the bound states are
bound states of D5-branes and D1-branes, the latter
being the heterotic string solitons of the type I theory
\edandjoe.  The mapping between instanton configurations
in four dimensions lifts to a map between M-brane and  
D-brane bound
states bound states in five dimensions.

To see how this might work, we can examine a
specific model which we can relate to the discussions
of type IIA physics in the previous section.
The heterotic dual to the local models we have been using
is a bit obscure (for example, one may use these local models
to ``engineer'' gauge groups of arbitrary rank,
while the rank of perturbative heterotic string
compactifications is limited.)  For the purposes
of this section we will choose as our
type IIA compactification manifold the
degree 24 hypersurface in
in $\IP^{4}_{1,1,2,8,12}$ (this example has
been worked out in detail in reference \asplect, which
also contains references to the original literature.)
This Calabi-Yau compactification contains the surface $\IF_{0}$
as a divisor at generic points in its space,
and is in fact an elliptic fibration over this Hirzebruch
surface. The local model of the previous section
can be thought of as a local description of the
$\IF_{0}$ divisor in this compact Calabi-Yau.
The model has three vector multiplets; two correspond to the
sizes of the two $\IP^{1}$s in $\IF_{0}$ and the last
corresponds to the size of the generic elliptic fiber.
The heterotic dual is the ${\rm E}_{8}\times {\rm E}_{8}$
string the gauge bundle on the $\bf K3$
has instanton number 12 in each ${\rm E}_{8}$ factor.
The vector multiplets correspond
to the geometry of the ${\bf T}^{2}$ ($T, U$) and the
heterotic dilaton $S$.

For this model, the map between the K\"{a}hler classes
of the Calabi-Yau compactification and the heterotic vector
moduli are straightforward and well-known.  As usual,
$S=1/g^{2}$ is the four-dimensional heterotic dilaton,
and $U$ and $T$ are the K\"{a}hler class and complex
structures of the ${\bf T}^{2}$ factor of the heterotic
compactification.
Let $T_{b}$, $T_{f}$, $T_{e}$ be the K\"{a}hler classes
of the base of the K3 fibration, the base of the generic
$\bf K3$ fiber, and the generic elliptic fiber respectively
(of course, for elliptic fibrations over $\IF_{0}$,
we can choose either $\IP^{1}$ to be the base of the
$\bf K3$ fibration \multiplekthree.)  The relation  
between these
moduli is
well known:
\eqn\modrel{
	\eqalign{
		T_{b} \propto S\cr
		T_{f} \propto (T-U)\cr
		T_{e} \propto U
	}
}
On the type IIA side, let us take $R_{11}\to\infty$ while
fixing the areas in the M-theory metric.  In this limit,
$$
	T_{b,f,e} \sim \sqrt{{{R_{11}}\over{\ell_{p,11}  } }},
$$
$T$ and $U$ both become large if one of the radii of the ${\bf
T}^{2}$ becomes
large.  $S$ becomes large at the same rate if we fix the
{\it five}-dimensional coupling to be finite.

In five dimensions the number of vector multiplets is
$h_{1,1}-1$ \cadavid.   The $\CU(1)$ corresponding
to the
K\"{a}hler class of the fiber on type IIA side
is $g_{\mu 6} + B_{\mu 6}$ on the heterotic side \antfertay
(where $6$ is the index along the $\bf S^{1}$).  
The symmetry will be enhanced
to $\su(2)$ when charged $W$ bosons become 
massless; these come from heterotic strings winding 
around the ${\bf  S}^{1}$ on the
heterotic side and two-branes wrapping the fiber $\IP^{1}$
on the type IIA side.  The ${\rm E}_{8}\times {\rm
E}_{8}$ string theory
in question also has a {\it heterotic} dual; the
dual string is the heterotic five-brane wrapped around
the $\bf K3$ \dmw.  For the dual pair we are
studying, this heterotic-heterotic duality is
mapped to fiber-base duality on the type IIA side,
if we think of the Calabi-Yau as being an elliptic
fibration of $\IF_{0}$ \multiplekthree.
If we wrap the heterotic fivebrane around the additional
${\bf S}^{1}$ of the five dimensional
compactification, it makes sense that it corresponds to
two-branes wrapped around the base $\IP^{1}$ in
the type IIA compactification.

In the type IIA compactification on $\bf K3$, we can
isolate the neighborhood of $\IF_{0}$ for the purposes
of computing the instanton
sum by making $T_{e}$ large. (On the heterotic side this would
correspond
to making $U$ and $T$ large but their difference finite.)
The claim is that the two-brane wrapped around the
homology
cycle $n[{\rm base}] + m[{\rm fiber}]$ is dual to
the bound state of $n$ fivebranes wrapped on
${\bf K3}\times {\bf S}^{1}$
and $m$ heterotic strings wrapped around ${\bf S}^{1}$.
Bound states
of these objects seem hard to compute, but in this case
there is a well-studied type $I$ dual, namely the
Gimon-Polchinski model \gipol.  The heterotic string
in this model is the D1-brane,
and the dual heterotic string is the D5-brane wrapped
around the $\bf K3$ \gipol\lotsofolk.
For proving type IIA-heterotic duality this
relationship is not very helpful as we would at best
trade one duality conjecture for another;  however if
we accept the type I-heterotic duality (or at least
accept the mapping between the relevant strings and branes),
the D-brane calculations might shed some light on the  
structure
of type IIA-heterotic duality.

The moduli space of the D1-brane/D5-brane system can
be reduced to an effective two-dimensional sigma-model,
{\it a l\'a}  \vafhag\bsvtop\vafinst\stromvaf.
In particular we can take the $\bf K3$ surface to be small
and study the effective theory on ${\bf S}^{1}$.
The D5-brane theory will be $\CU(n)$ for $n$ D5-branes
\gipol\lotsofolk.  The D1-branes should appear as
instantons \dougone\vafinst\dougtwo
on the part of the D5-brane which is wrapped around the
$\bf K3$. We also need to understand the worldsheet
supersymmetry. The states should be BPS states which  
break half of
the
supersymmetries of the heterotic theory (because the dual
M-theory two-brane configurations do.)  Following the
discussion in \jeffandy, we start with the heterotic string
in ten dimensions.  It has one supercharge transforming
as a {\bf 16} of $\so(9,1)$.  This representation
decomposes into the representation
\eqn\sixtofiverep{
	{\bf 16} \to ((\half,0),(\half,0))^{+}
		+ ((\half,0),(0,\half))^{-}
		+ ((0,\half),(\half,0))^{-}
		+ ((0,\half),(0,\half))^{+}
}
of the group
$$
	\so(4)\times\so(4)\times\so(1,1)
$$
(the superscript denotes the $\so(1,1)$ chirality.)
The second 
\hbox{$\so(4)=\su(2)_{1}\times\su(2)_{2}$} is
the spatial rotation group of the uncompactified directions.
Let the $\bf K3$ break the supersymmetries transforming under
$(1/2,0)$ in the first $\so(4)$ factor.  The fivebrane
wrapped around $\bf K3$ is the dual heterotic string;
it leads to a  BPS state breaking half of the spacetime
supersymmetries.  Let it break supersymmetries
transforming under $(1/2,0)$ of the second $\so(4)$
factor (i.e. transforming as a doublet of $\su(2)_{1}$.)
The fermion zero modes of the D1-brane have a single
chirality (as the D1-brane of type I string theory is
supposed to be the heterotic string)\edandjoe,
and before introducing the $\bf K3$ and the D5-brane
we expect eight of them.
We can orient the D1-brane so that the zero-modes --
and thus the supersymmetries broken by the D1-brane --
have negative chirality with respect to $\so(1,1)$.
The supersymmetries
unbroken by the $\bf K3$ but broken by the fivebrane and
onebrane lead to fermion
zero modes transforming as $((0,1/2),(1/2,0))^{-}$,
which will become left-moving superpartners of the
translational zero-modes. The unbroken
supersymmetries transform as
$((0,1/2),(0,1/2))^{+}$, so we
have a $(4,0)$ sigma-model.  The $\su(2)$ R-symmetry
is the factor of the spatial rotations in $4+1$ dimensions
under which the unbroken supersymmetries transform.

It would be nice to be able to show that the contributions
of these bound states to the prepotential of the $\su(2)$
gauge theory on the heterotic side match the contributions
we have discussed on the M-theory side.  The comparison
is similar in spirit to the calculations in
\kmvbps\lmwnoncrit\mnwnoncrit.  In particular,
for M D1 branes and $n$ D5-branes wrapped around
the circle, we would expect that for fixed $n$ and large
M, the contribution should be
$$
\gamma_{n}m^{4n-3}
$$
as it is on the M-theory side \kkv.  Unfortunately we
have not been able to reproduce this result at
present, but we will instead present our speculations
as to how the calculation should be set up.

As in our study of M2-branes, the broken supersymmetry should
form a half-hypermultiplet transforming 
as $2(0,0)\oplus(0,1/2)$
under the five-dimensional massive little group.  Because
the bound states we are interested are BPS states, the
left-moving
sector must be in the Ramond ground state.  This ground
state can be decomposed into representations of the $\su(2)$
R-symmetry, which is the same as the spin under $\su(2)_{1}$.
As before, for each such representation with spin $k/2$,
the contribution to the prepotential for the one-loop
contribution of this bound state should be proportional
to
$$
	(-1)^{k}(k + 1)\ .
$$
Generally we must also include contributions of the
right-movers.
Because the heterotic strings and their duals correspond to
$W$-bosons, they must each have one unit of momentum
around ${\bf S}^{1}$.
It seems, then, that the quantity to calculate is related to
the partition function
$$
{\rm Tr}\left\{
(-1)^{F}\bar{q}^{\bar{L}_{0}}q^{L_{0}}\right\}\ ,
$$
projected onto $\bar{L}_{0}=0$, $L_{0}=m+n$.  Note that
the sigma model is somewhat more complicated than the
usual effective sigma model for D1-D5 bound states in
type I models.
In particular, the Gimon-Polchinski model contains
D5-branes (let us label them as D5$^{\prime}$ branes)
filling the space transverse to the $\bf K3$, as well as
the usual
D9-branes.  So there will be $15^{\prime}$ and $55^{\prime}$
strings,
and we expect the target space of this effective sigma model
to be somewhat more complicated than the usual instanton
moduli space.

\newsec{Conclusions}

The principal lesson of this paper is that if we
isolate a developing singularity in a Calabi-Yau
manifold, it still makes sense to claim that
we have isolated the gauge theory dynamics if
we understand the gauge theory as living on
$\IR^{4}\times {\bf S}^{1}$.  In addition, it appears
that this point of view may be helpful in understanding the
relation between spacetime and worldsheet nonperturbative
effects both within the type IIA framework
and in the relation between type IIA and heterotic
compactifications.
One surprise is that the full four-dimensional instanton
sum is {\it perturbative} from the point of view of
the five-dimensional gauge theory.

There is clearly some unfinished work in this paper.  First,
we need to understand the absence of the linear terms
in the gauge coupling that promote $\log (1-e^{-t})$ to
$\log \sinh (t/2)$.  One issue is that this term
has a natural interpretation in the M-theory picture
as being the loop effect of a wrapped M2-brane in
five uncompactified dimensions, but there
is no obvious (geometrical) map
to a string theory process as there
is for the instanton sum.  Secondly, we need to
understand better how to treat the K\"{a}hler classes
that are missing from the discussion in section 3.
Finally, it would be nice to show that the string-fivebrane
bound states we have discussed give the same contribution
to the prepotential of the heterotic theory as the dual
two-brane states give to the prepotential of the type IIA  
theory.

One of the original motivations of this work was to calculate
the worldsheet instanton sum directly for toric varieties
using localization techniques described by Kontsevich \kenu.
Since then, Givental \given\giventwo\ claims to have
proven a piece of mirror symmetry (at least, the resulting
expressions for the prepotential and the mirror map), using
a variant of these techniques to calculate the equivariant
Gromov-Witten invariants.  It would be nice
to translate this proof into something a bit more physical
\wepromise.  Our hope is that this will
not be a completely academic exercise.  For example,
some of the structure of the calculation
outlined by Kontsevich (i.e. the graphical expansion)
might have a direct interpretation in terms of
D-brane states.\foot{This possibility was suggested to
us by C. Vafa}  We might also hope that this structure would
appear in the heterotic dual, in particular in the bound state
problem we outlined in section 4.

We can imagine arguing that the
D-brane bound state calculation of section 4 can be deduced
entirely from gauge theory considerations.  The argument
might go roughly as follows.  Each term in the
instanton sum of a four-dimensional gauge theory will
have a coefficient arising from the integration over the  
instanton moduli space. This space is described by
ADHM data \adhm (see also \christ).  
This data is the moduli space of
vacua of an $\CN=1$, $d=6$ gauge theory 
\smallinst\dougtwo; for $\su(2)$ 
the $k$-instanton
sector is described by a $\CU(k)$ gauge theory.   
We could
imagine then that the coefficients of the spacetime instanton
expansion are related to correlation functions of this  
gauge theory
(or an appropriate twisting.)  We might further
argue that for supersymmetric spacetime theories,
the R-symmetry acts as an unbroken Lorentz symmetry
of the auxiliary $d=6$ theory (and not just on the moduli
space of vacua).  Thus the $\su(4)\cong\so(6)$
part of the
R-symmetry group acts as the Lorentz symmetry of
the $d=6$ theory on ${\bf T}^{6}$ or $\IR^{6}$,
while the $\su(2)$ R-symmetry acts as the 
part of the Lorentz group surviving the $\bf K3$
compactification.
The correlation functions would
also be broken up into instanton sectors of this gauge theory.
Of course, all of is natural from the D-brane point of  
view; we
have only tried to abstract the features of the D-brane  
calculation
which arise naturally when thinking only about gauge theories,
simply by symmetry considerations.
It is not clear from gauge theory considerations  
what the
relevant correlation functions are.  Some progress has been
made in integrating over the ADHM moduli space (and over
hyper-K\"{a}hler quotients in general) \mnspromise.  One  
interesting
feature is that the integrand has a natural $\CU(1)$  
action and
thus one can apply a localization formula to the integral.
Perhaps this provides a link to the type IIA worldsheet  
instanton
calculation outlined by Kontsevich.

It would be interesting to relate this work to that in  
\hm\hmalg.
We have described a calculation which is related to brane
bound states on both the heterotic and type IIA sides of a dual
pair.  These bound states become massless (classically)
at a boundary of the compactification moduli space and
may reflect some large symmetry algebra.  In addition,
our rederivation of the Gromov-Witten invariants indicates
that the worldsheet instanton moduli space should
be the same as the moduli space of D-branes, namely
the space of coherent sheaves as conjectured in \hmalg.
This puts a little meat on a conjecture of Kontsevich
regarding a generalized formulation of mirror symmetry
\konthom.

Finally, we know that both the $E_{8}\times E_{8}$
heterotic string and the type IIA string are naturally
described from the point of view of
M-theory.  It would be interesting to understand if this
duality, and in particular the relation between the bound  
states
that we have described, arises from a fundamental
duality symmetry of M-theory compactifications.
The matrix model of M-theory \bfss\ might help.
M-theory compactifications on Calabi-Yau manifolds
are not yet understood; however,
the authors of \berkroz\ have claimed that
the proposal outlined in \rozali\ and \berkrozseib\ makes
the duality between the heterotic string on $T^{3}$ and
M-theory on $\bf K3$ manifest.  A microscopic
understanding of heterotic-type IIA duality might
lie in this direction.

\vskip 1.5cm
{\bf Acknowledgments}

We would like to thank
P.~Aspinwall, T.~Banks, O.~Bergman, A.~ Gerasimov, A.~ Gorsky, 
A.~Hashimoto, D.~Kabat,
S.~Kachru, A.~ Losev,  J.~Maldacena, G.~Moore, A.~ Morozov, 
N.~Seiberg, S.~Shatashvili, E.~Silverstein, C.~Vafa, E.~Witten and
B.~Zwiebach for
discussions, explanations, comments, and encouragement.
A.~L.~ would like to thank the Rutgers high energy theory
group for
their hospitality while much of this paper was being
written.  He would also like to thank G. Zuckerman
and the Yale mathematics department for their hospitality
while a draft was being completed.
N.~ N.~ would like to thank UNAM at Mexico City and
especially Prof.
A.~ Turbiner for hospitality while part of
this work was done.

The research of A.~L.~
was supported in part by NSF grant
PHY-92-18167; the research of N.~N.~  
was supported in part by the Harvard Society of Fellows,
by NSF grant PHY-92-18167, and by RFFI grants
96-02-18046 and 96-15-96455.

\listrefs
\end